\documentclass[a4paper,twoside]{article}
\newcommand{\affil}[1]{$^{\rm #1}$}
\usepackage[authoryear]{natbib}
\bibpunct{(}{)}{;}{a}{}{,}

\usepackage{graphicx}
\usepackage{url}
\usepackage{fixltx2e}
\date{} %Please leave the date blank
\newcommand{\mab}{\mbox{$m_{AB}$}}

\newcommand{\degsq}{\mbox{deg$^{2}$}}
\newcommand{\degs}{\mbox{$^{\circ}$}}
\newcommand{\msun}{\mbox{M$_{\odot}$}}

\newcommand{\si}{\mbox{$\sim$}}
\newcommand{\mm}{\mbox{$\mu$m}}
\newcommand{\mj}{\mbox{$\mu$Jy}}

\newcommand{\vb}{\emph{V}-band}
\newcommand{\ib}{\mbox{\emph{i}-band}}

\newcommand{\Rb}{\mbox{\emph{R}-band}}
\newcommand{\jb}{\emph{J}-band}
\newcommand{\hb}{\emph{H}-band}
\newcommand{\kb}{\emph{K}-band}
\newcommand{\kd}{\mbox{\emph{K}$_{d}$}}

\newcommand{\kdb}{\mbox{\emph{K}$_{d}$-band}}

\title{\large\bf\flushleft The Science Case for PILOT II: the Distant Universe}
\author{\parbox{\textwidth}{\flushleft
\vspace{-0.5cm}
{\it J.S.~Lawrence\affil{\,A,J}, M.C.B.~Ashley\affil{\,A}, A.~Bunker\affil{\,B}, R.~Bouwens\affil{\,C}, D.~Burgarella\affil{\,D},
M.G.~Burton\affil{\,A}, N.~Gehrels\affil{\,E}, K.~Glazebrook\affil{\,F}, K.~Pimbblet\affil{\,G}, R.~Quimby\affil{\,H}, W.~Saunders\affil{\,B},
J.W.V.~Storey\affil{\,A}, J.C.~Wheeler\affil{\,I}}\\
\vspace{0.4cm}
{\small \affil{A}\,School of Physics, University of New South Wales, NSW 2052, Australia}\\
{\small \affil{B}\,Anglo-Australian Observatory, NSW 1710, Australia}\\
{\small \affil{C}\,Department of Astronomy and Astrophysics, University of California Santa Cruz, Santa Cruz, CA 95064, USA}\\
{\small \affil{D}\,Observatoire Astronomique de Marseille Provence, Universit\'{e} d$'$Aix-Marseille, Marseille 13388, France}\\
{\small \affil{E}\,NASA/Goddard Space Flight Center, Greenbelt, MD 20771, USA}\\
{\small \affil{F}\,Centre for Astrophysics and Supercomputing, Swinburne University of Technology, Hawthorn, VIC 3122, Australia}\\
{\small \affil{G}\,Department of Physics, University of Queensland, Brisbane, QLD 4072, Australia}\\
{\small \affil{H}\,Astronomy Department, California Institute of Technology, Pasadena, CA 91125, USA}\\
{\small \affil{I}\,Department of Astronomy, University of Texas, Austin, TX 78712, USA}\\
{\small \affil{J}\,JSL now at Department of Physics and Electronic Engineering, Macquarie University, NSW 2109, Australia; and Anglo-Australian Observatory, NSW 1710, Australia; Email: jsl@physics.mq.edu.au}}}
\begin{document}
\twocolumn[
\begin{changemargin}{.8cm}{.5cm}
\begin{minipage}{.9\textwidth}
\vspace{-1cm}
\maketitle
%
%
%%%%%%%%%%%%%     ABSTRACT    %%%%%%%%%%%%%
\small{\bf Abstract:} PILOT (the Pathfinder for an International Large Optical Telescope)
is a proposed 2.5~m optical/infrared telescope to be located at Dome~C on the Antarctic
plateau. The atmospheric conditions at Dome~C deliver a high sensitivity, high
photometric precision, wide-field, high spatial resolution, and high-cadence imaging
capability to the PILOT telescope. These capabilities enable a unique scientific
potential for PILOT, which is addressed in this series of papers. The current paper
presents a series of projects dealing with the distant (redshift $>1$) Universe, that
have been identified as key science drivers for the PILOT facility. The potential for
PILOT to detect the first populations of stars to form in the early Universe, via
infrared projects searching for pair-instability supernovae and gamma-ray burst
afterglows, is investigated. Two projects are proposed to examine the assembly and
evolution of structure in the Universe: an infrared survey searching for the first
evolved galaxies at high redshift, and an optical survey aimed at characterising
moderate-redshift galaxy clusters. Finally, a large-area weak-lensing survey and a
program to obtain supernovae infrared light-curves are proposed to examine the nature and
evolution of dark energy and dark matter.

%%%%%%%%%%%%%     KEYWORDS    %%%%%%%%%%%%%

% and available from http://www.journals.uchicago.edu/ApJ/keywords_text.html.

%%%%%%%% DO NOT EDIT %%%%%%%%%%%%
\medskip
\medskip
\end{minipage}
\end{changemargin}
]
\small
%%%%%%%% EDIT FROM HERE %%%%%%%%%%%%

\section{Introduction}

PILOT (Pathfinder for an International Large Optical Telescope) is proposed as a high
spatial resolution wide-field telescope with an optical design and an instrument suite
that are matched to the Dome~C atmospheric conditions. These conditions have been shown
to offer high infrared sensitivity, due to the low atmospheric thermal emission and low
water-vapour column-density \citep{Lawrence_04,Walden_e_05,Tomasi_e_06}; high spatial
resolution and high photometric precision, due to the unique atmospheric turbulence
structure above the site \citep{Lawrence_e_04,Agabi_e_06,Kenyon_e_06,Trinquet_e_08}; and
a high cadence, due to the high latitude of the site and the high percentage of
cloud-free conditions \citep{Kenyon_Storey_06,Mosser_Aristidi_07}.

The scientific justification for the PILOT telescope has evolved from earlier work
\citep{Burton_e_94,Burton_e_01,Burton_e_05} in parallel with the telescope and instrument
suite design \citep{Saunders_e_08a,Saunders_e_08b}. The PILOT science case is presented
here in a series of three papers. Paper I \citep{Lawrence_e_09a} gives a summary of the
science case, and an overview of the project (including the telescope design, expected
performance, and observing strategies). The current paper (Paper II) presents a series of
science projects for the PILOT facility that are aimed at observing and understanding the
distant Universe (i.e., beyond a redshift of $z\approx1$). Paper III
\citep{Lawrence_e_09b} discusses PILOT science projects dealing with the nearby Universe
(i.e., the Solar System, the Milky Way, and nearby galaxies).

The baseline optical design for PILOT, described in \citet{Saunders_e_08a}, comprises a
2.5~m Ritchey-Chretien telescope with an f/10 overall focal ratio. The telescope is
housed in a temperature- and humidity-controlled dome that is mounted on top of a \si30 m
high tower in order to elevate the main mirror above the majority of the intense
ground-layer turbulence. A fast tip-tilt secondary mirror is used for guiding and to
remove residual boundary-layer turbulence and tower wind-shake. As detailed in Paper I
\citep[see also][]{Saunders_e_08b} the baseline instrument suite for PILOT includes the
PILOT VISible Camera (PVISC), a wide-field ($40'\times40'$) optical imaging camera with a
spatial resolution of \si0.3$''$ over the wavelength range 0.4--1~\mm; the PILOT
Near-InfraRed Camera (PNIRC), a wide-field ($10'\times10'$) near-infrared camera
achieving a similar spatial resolution over the wavelength range 1--5~\mm; the PILOT
Mid-InfraRed Imaging Spectrometer (PMIRIS), a wide-field ($40'\times40'$) mid-infrared
instrument operating from 7--40~\mm, with several spectral-resolution modes; and the
PILOT Lucky Imaging Camera (PLIC), a fast optical camera for diffraction-limited imaging
over relatively small-fields ($0.5'\times0.5'$) in the visible.

The ``distant Universe" science cases described in this paper are divided into three key
themes: \emph{first light in the Universe}, \emph{the assembly of structure}, and
\emph{dark matter and dark energy}. The projects proposed under these themes have been
identified to take advantage of the unique discovery space of the PILOT telescope,
enabled by the Dome~C site conditions and the telescope design. For each project, the
context and impact is discussed, along with the capabilities of competing (and
synergistic) facilities. Additionally, the observational requirements are identified. It
is shown that the observational requirements for these projects can either only be
achieved with PILOT or can be achieved with other facilities but with much lower
efficiency. While many of the science projects described here (and in Paper III) require
large amounts of observing time (several seasons in some cases), the observing strategies
proposed in Paper I should allow the majority of these science projects to be completed
within the proposed ten-year lifetime of the PILOT facility.

\emph{First light in the Universe} (Section 2) projects are linked by the desire to
detect and characterise the earliest stellar populations to form. Here, we investigate
the potential for PILOT to observe the signatures of the final evolutionary stages of
very high-redshift (i.e., $z>10$) Population III super-massive stars. A deep wide-area
infrared survey should allow the detection of pair-instability supernovae. A
target-of-opportunity program of near- to mid-infrared satellite-alert follow-up should
allow the detection of gamma-ray burst afterglows.

\emph{The assembly of structure} theme (Section 3) links studies of distant galaxies and
galaxy clusters. We investigate the potential for PILOT to detect distant galaxies via
deep, wide-field, infrared surveys at \kdb. In particular, the first populations of
evolved galaxies to form, in the redshift range $z = 5$--7, should be identifiable with
PILOT, when combined with other telescope survey data, via photometric observations
around the redshifted Balmer break wavelength. We also propose an optical investigation
of a significant number of X-ray selected galaxy clusters in the redshift range
$z=0.6$--1.5. This study will further our understanding of cluster growth, structure, and
evolution.

The \emph{dark matter and dark energy} theme (Section 4) involves two projects that aim
to investigate the evolution of cosmological parameters. Firstly, a very wide-area
optical sky survey is proposed to measure large-scale structure via the weak-lensing
signature of distant galaxies. From this structure, constraints on the nature and
evolution of dark matter and dark energy can be derived. Secondly, a PILOT survey aimed
at obtaining large numbers of supernovae Ia light-curves at near-infrared wavelengths,
where they are largely unaffected by dust extinction and reddening, will play an
important role in removing (or confirming) concerns that dust may affect current
constraints placed on cosmological parameters from optical data.

%%%%%%%%%%%%%%%%%%%%%%%%%%%%%%%%%%%%%%%%%%%%%%%%%%%
\section{First Light in the Universe}
%%-----------------------------------------------------
\subsection{Pair-Instability Supernovae at \\High Redshift}
\subsubsection{Impact}

Pair-instability Supernovae (PISNe) are predicted to be extremely powerful thermonuclear
explosions that result from the softening of the equation of state by electron/positron
pair formation and the subsequent contraction, heating, ignition, and total disruption of the oxygen cores of especially massive stars. Current theoretical estimates of the
progenitor mass range and resulting explosion energy are 140 to 250~\msun\ and $2\times
10^{51}$ to $100\times10^{51}$ ergs, respectively \citep{Heger_Woosley_02}. Conditions in the early Universe at high redshift are predicted to be especially conducive to the
formation of such massive stars \citep{Bromm_Larson_04}. The existence and distribution
with redshift of PISNe may thus provide a powerful way of probing the history of chemical enrichment and reionisation and of the physical conditions in the environments around the first structures to form in the Universe.

The pair-instability process is predicted to occur for massive stars in very low
metallicity environments. The metal-free environment lowers the fragmentation of
collapsing clouds, therefore resulting in larger protostellar cores
\citep[e.g.,][]{Abel_e_02}. It also increases the accretion rates that are not arrested
by the processes of radiation pressure, bipolar outflows, and rotation. Finally, it
suppresses the efficiency of line-driven wind mass loss so that massive stars do not shed their hydrogen envelopes before core He burning is ended \citep{Kudritzki_02}. These
factors lead to the large stellar masses essential for the pair-instability process. The
low metallicities required suggest that PISNe should be numerous amongst the first
(Population III) stars.

For stars in the appropriate mass range, the pair-instability process involves the
complete disruption of the progenitor. In this case the whole core of the star
(20--60~\msun) is converted to radioactive $^{56}$Ni, creating an object that is
intensely bright, typically $-22$ absolute mag, and lasting for typically 50--150 days.
For stars of mass lower than \si140~\msun\ the nuclear burning initiated by the pair
instability is not sufficient to halt the collapse; such objects thus leave behind a
black hole remnant \citep{Heger_Woosley_02}. Stars of initial mass greater than
\si250~\msun\ are thought to collapse via photo-disintegration, also leaving a blackhole
remnant \citep{Heger_Woosley_02}.

As the PISN process involves the complete disruption of the progenitor star, and since
they should be numerous amongst Population III stars formed in the range
$z\approx10$--20, this process is thought to be primarily responsible for enriching the
IGM metallicity above the critical level sufficient to halt very massive star formation
and allow Population II stars to form \citep[e.g.,][]{Scannapieco_e_03}. However, metal
enrichment is a local process that is not expected to raise metallicity in a completely
homogeneous way. For this reason, it is expected that PISN should occur at much lower
redshifts (to a least $z=6$) in isolated regions of gas where metals have not yet
propagated \citep{Mackey_e_03}. There is, as yet, no sign of the specific chemical
signature of PISNe in abundance studies of low metallicity stars, but this may be due to
selection effects \citep{Karlsson_e_08}.

\begin{figure}[t]
\begin{center}
\includegraphics[width=7.5cm]{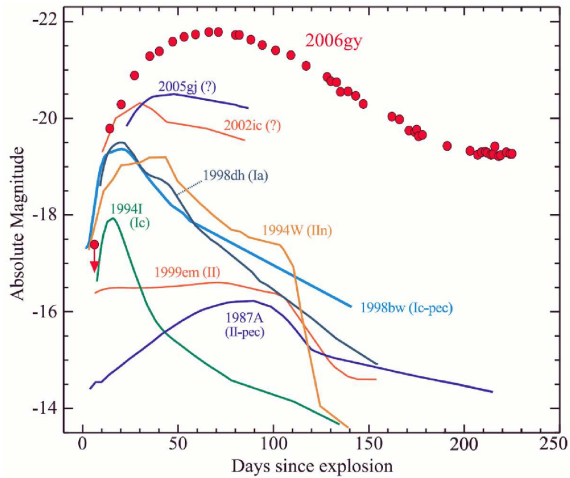}
\caption{Absolute \Rb\ light-curve for SN2006gy compared with those of other SNe (types
indicated in figure). From \citet{Smith_e_07}. }\label{PISN_SN2006}
\end{center}
\end{figure}

Models predict that PISNe will be observationally distinct from other SNe since they are
much more luminous, exhibit a much slower rise time to peak luminosity, and stay brighter
for a longer time. SN 2006gy, the light curve of which is shown in
Figure~\ref{PISN_SN2006}, is a recently discovered local (within \si100 Mpc) SN which
meets these characteristics \citep{Smith_e_07}. This SN took \si70 days to reach peak
luminosity, compared with \si20 days typical for other SN types, and reached a peak
absolute brightness of $-22$ magnitudes, a factor 2--6 times brighter than other SN
types. \citet{Woosley_e_07} suggested that this object could be a pulsational type of
PISN. More recent work, however, suggests alternative explanations
\citep{Smith_McCray_07,Smith_e_08}.

\subsubsection{Observing Strategies and Detection Rates}
No confirmed detection of a PISN has yet occurred. As these objects are theoretical
predications based on extrapolations from existing data and models, there is a large
uncertainty as to their expected characteristics. A wide range of expected number
densities of PISNe have been reported. Values range from 1 to 36~objects per \degsq\ per
year per unit redshift at $z=6$, and from 0.3 to 6~objects per \degsq\ per year per unit
redshift at $z=15$ \citep{Scannapieco_e_05,Weinmann_Lilly_05,Wise_Abel_05}. This large
range arises from uncertainties in several of the model input parameters, primarily the
fraction of gas per primordial object converted into stars (i.e., the star formation
rate, SFR), the number of PISNe per unit mass of stars formed, and the mass distribution
for Population III stars (i.e., the initial mass function, IMF). There is a also a large
range of predicted values for the PISN rest-frame ultraviolet time interval, specified as the time the event is one magnitude brighter than the peak brightness, from a few days
\citep{Heger_e_02,Weinmann_Lilly_05} to \si30~days \citep{Scannapieco_e_05} to
\si100~days \citep{Wise_Abel_05}. This corresponds to an observed-frame time interval
ranging from a few tens of days at low redshift to more than a year for very
high-redshift objects. Finally, there is a range of expected absolute peak brightness for these objects, ranging from $-$20 to $-$24 for progenitor masses of 250~\msun\
\citep{Scannapieco_e_05,Weinmann_Lilly_05,Wise_Abel_05} arising because of the range of
simplifying assumptions used in the models. These uncertainties will have significant
implications for the observing strategy and required cadence for any PISN search. If the
values for the time interval are at the upper end of the expected range, then multi-year
programs will be essential. If the values for the peak brightness are at the lower end of the expected range, then very long integrations will be required per field.

For relatively low-redshift PISNe, a search in \ib, above the limit for Lyman~$\alpha$
absorption at $z\approx5$, is appropriate. The required sensitivity for detection is
$\mab\approx27$ \citep{Scannapieco_e_05}. Such a search would require an integration time per field of \si1.8~hours for PILOT's PVISC instrument. For a dedicated survey (i.e.,
100\% of telescope time spent on this project), a 3000~hour search (i.e., one winter
season) would allow over 800~\degsq\ to be covered. This would be expected to find
thousands of PISN in the range $z=2$--3 and a few tens of PISN out to $z=4$
\citep{Scannapieco_e_05}, assuming that these objects exist at such low redshifts. Other
ground-based wide-area optical survey facilities such as LSST and Pan-STARRS-4 should be
capable of similar detection rates at this wavelength, but only if they spend much
longer, and go much deeper, than their currently planned survey strategies.

\begin{figure}[t!]
\begin{center}
\includegraphics[width=7.5cm]{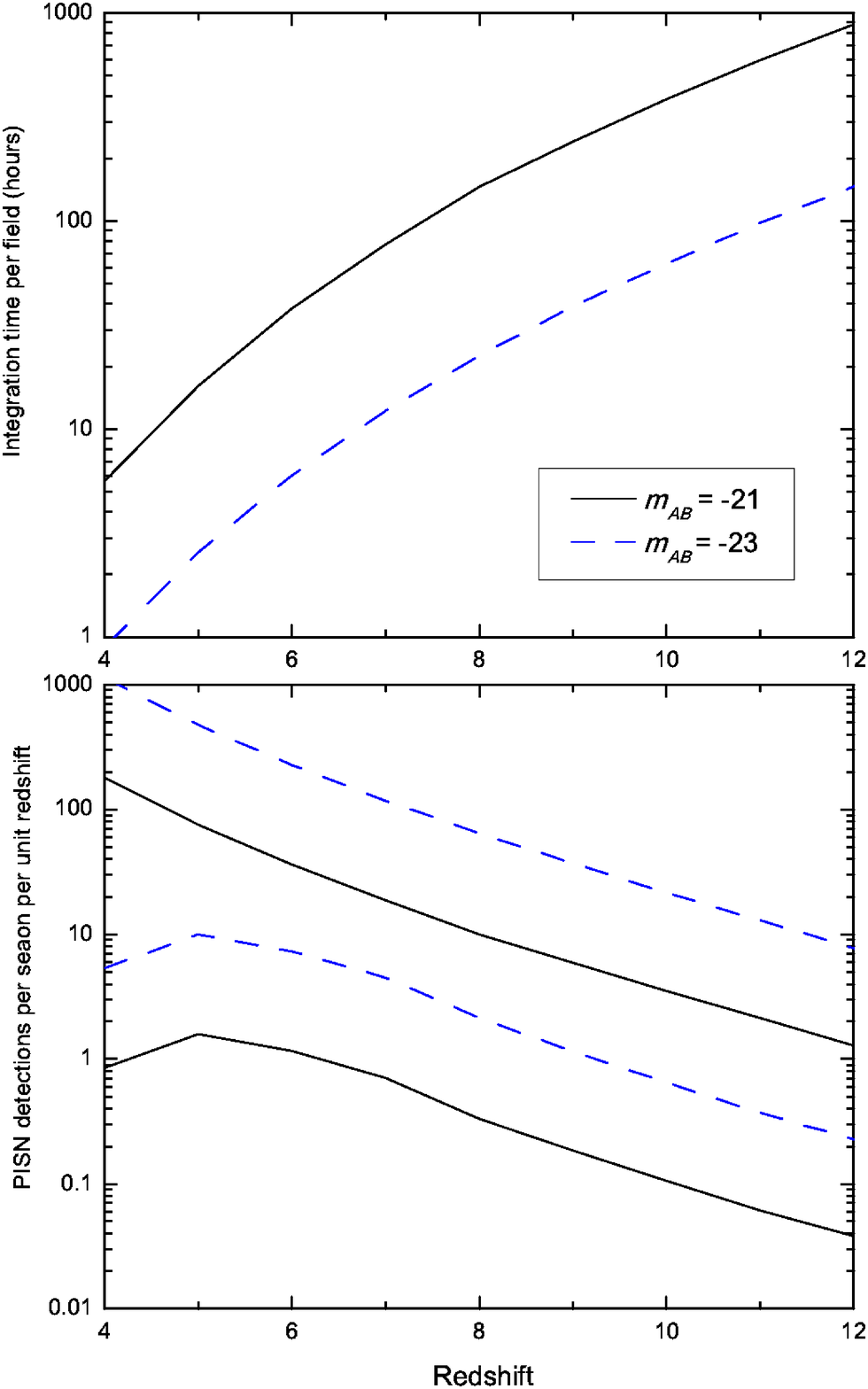}
\caption{Top: integration time required per field for PISN detection 0.5 magnitudes
below peak brightness at \kdb\ with PNIRC for PISN absolute brightness of
$\mab=-21$ and $-23$. Bottom: expected number of PISNe detected by PILOT per
winter season (3000 hours) as a function of redshift for a dedicated \kdb\ survey,
for PISN absolute brightness of $\mab=-21$ and $-23$. In each case,
the upper and lower detection limits are based on PISN number density
estimates from \citet{Scannapieco_e_05} and \citet{Weinmann_Lilly_05}.
A rest-frame ultraviolet time interval of 20~days, and a search pattern that repeats
at the corresponding observed-frame time interval, is assumed.
Multi-year searches will thus be required to find very high-redshift objects. }\label{PISN_rate}
\end{center}
\end{figure}

A more promising opportunity for PILOT is to search for higher-redshift PISNe in the
near-infrared \kdb\ with PNIRC. Figure~\ref{PISN_rate} shows the required integration
time for PISNe with an absolute brightness ranging from $\mab=-21$ to $-23$, and the
expected detection rate per season as a function of redshift. For PISNe at the upper end
of the expected range of brightness, integration times per field range from 1~hour at
$z=4$, to 50~hours at $z=10$, corresponding to total search areas of 3--90~\degsq\ (i.e., before repeat visits are required). Using pessimistic surface density predictions, PILOT
should detect at least one PISN at $z\approx10$ per season. Optimistic models allow the
detection, per season, of several hundred PISN in the range $z=6$--10, and at least one
beyond $z=12$ (for very long integrations). For PISN at the lower end of the expected
range of brightness, much longer integration times are required, resulting in very small
search areas and a low probability of detecting any PISN at redshifts greater than about
$z=8$. SN2006gy type objects (with peak brightness of $\mab\approx-22$) should be
discovered at least out to $z=6$, and possibly hundreds could be discovered per season.
While there are large uncertainties associated with these predictions, they illustrate
the potential for PILOT in this area.

Other transient phenomena will also be detected by the PISN survey, and while these may
comprise an interesting sample in their own right, they represent contaminates to the
PISN sample that must be identified and excluded. This can effectively be done by
breaking up the long integrations into multiple epochs and with supporting multi-band
observations of the host galaxies. It is difficult or impossible to differentiate between a high-redshift PISN and a low-redshift transient, such as a Type Ia SN, given only a
single detection from a very long \kd\ exposure; however, if the integration were broken
up over several weeks, then each epoch would sample a photometrically distinct portion of the Type Ia supernova's light curve, but the slowly evolving PISN would remain
essentially constant. Similarly, AGN exhibit a red-noise power spectral distribution and
may be seen to vary across the multiple epochs. However, some Type II SNe can show nearly constant photometric plateaus, and $z=1.5$ Type II-P supernova could have the same peak
apparent brightness as a $z=10$ PISN. Such contaminants may be further weeded out by
comparison with deep template images. An \Rb\ detection of the host, for example, would
place the transient at $z<5$. If the template images extended into multiple bands, a
photometric redshift could further refine the distance. Finally, multi-epoch high-resolution spectroscopic follow-up of any PILOT PISN detection using 8-m class mid-latitude telescopes (and/or JWST) will help refine redshift estimates for these objects.

\subsubsection{Other Facilities}
The faintness of PISNe in the infrared and their inherent rarity calls for a very
specialised facility to detect them. As indicated here, it requires a telescope with a
wide field-of-view and a high infrared sensitivity. This is a good match to the
capabilities of PILOT. The expected rarity of these events demonstrates the
importance of field-of-view. This science case would be significantly strengthened, for
example, by increasing the field-of-view of PNIRC by a factor of 4 (i.e., using a larger
mosaic of detectors).

The \kdb\ is chosen here to illustrate the potential for PILOT to detect high-redshift PISNe as this is the most efficient wavelength to search in the near-infrared (PILOT should observe to a depth of 0.3 and 0.7 AB magnitudes fainter at \kdb\ than at \jb\ and \hb\ respectively). The \kdb\ is also where PILOT has the largest advantage compared to mid-latitude telescopes; PILOT is a factor of eight times faster to reach a given depth over a given area at this wavelength than VISTA, for example. For a mid-latitude telescope, however, it is more efficient to observe at \jb\ or \hb\ than at \kb\ and PILOT is only marginally more sensitive at these lower wavelengths. The real benefit to a PISN search at \kdb\ is thus the possibility to detect higher redshift objects that cannot be observed at lower wavelengths. \jb\ and \hb\ correspond to Lyman~$\alpha$ redshifted to $z=10$ and $z=13$ respectively. Peak spectral emission for PISN is expected to occur at wavelengths longer than Lyman~$\alpha$ \citep{Scannapieco_e_05}. Mid-latitude ground-based telescopes are thus precluded from detecting significant numbers of these objects at redshifts beyond about $z=6-8$.

While JWST is much more sensitive in the thermal infrared than PILOT (it should reach
$\mab= 28$ at 2.5~\mm\ for $R=10$ in \si13~minutes\footnote{Based on scaling the quoted
JWST sensitivity limits at \url{http://www.stsci.edu/jwst/science/sensitivity/}}), the
infrared camera NIRCam has a smaller field-of-view than PNIRC. Additionally, the time
associated with slewing and settling precludes JWST from conducting surveys over large
(degree scale) regions of sky. PILOT will thus probe a different region of the possible
PISN parameter space. If PISNe occur at the faint end of the expected brightness range
and are common, then only JWST will find them at high redshift \citep{Gardner_e_06}. If
PISNe occur at the bright end of the expected brightness range with low surface density,
then JWST will not be able to cover a sufficient area of sky and PILOT will be the only
facility capable of detecting them (JWST will then be a useful facility for following up any such PILOT detections). If these objects, however, are both very faint and very rare, then detection of high-redshift PISNe will not be possible with any currently
envisaged telescope.

%%-----------------------------------------------------
\subsection{High-Redshift Gamma-Ray Bursts}
\subsubsection{Impact}
Gamma-ray bursts (GRBs) are extremely powerful explosions, with total energies ranging
from $3\times10^{51}$ to $2\times10^{54}$~ergs. They are rare astrophysical events
believed to be produced by internal shocks taking place in a collimated jet of
relativistic particles that is emitted by a central engine. Satellite observations have
demonstrated a bimodal temporal distribution to this high-energy emission
\citep{Kouveliotou_e_93}. Short ``hard" bursts (lasting less than \si2~seconds) have been attributed to the coalescence of binary compact objects in nearby galaxies
\citep{Fox_e_05,Gehrels_e_05}. Long-duration bursts (lasting less than a few minutes)
have been associated with the gravitational core-collapse of massive stars in distant
galaxies \citep{MacFadyen_Woosley_99}. In both cases the interaction of the jet with the
surrounding medium produces an afterglow of much longer duration, typically lasting hours to days \citep{Akelof_e_99,Nousek_e_06,Zhang_e_06}. This afterglow, observed at optical,
X-ray, and radio wavelengths, is thought to be less directional than the high-energy
(gamma-ray) emission. It is thus expected that geometries will exists that result in the
detection of an ``orphan" afterglow (i.e., unaccompanied by gamma-ray emission).
Approximately 40\% of GRBs are known to be ``dark bursts", i.e., they have no optical
counterpart to faint limits. This is believed to occur for cases of high obscuration,
where the optical emission is attenuated, and high redshift, where the optical emission
falls at a shorter wavelength than the Lyman~$\alpha$ cut-off in the source frame.

Due to their high intrinsic luminosity, gamma-ray bursts offer a powerful probe of the
Universe at a range of cosmological distances. Both the host-galaxy interstellar medium
and the intergalactic medium are imprinted on the spectrophotometric signature of the
gamma-ray burst optical/infrared afterglow \citep{Prochaska_e_07}. Their association with the final stages of the evolution of massive stars provides information on the star
formation history of the Universe at different cosmological eras \citep{Jakobsson_e_05}.
The highest redshift GRB so far detected is at $z = 6.3$ \citep{Kawai_e_06}. Theories
suggest that long duration GRBs should be numerous at higher-redshifts in the range $z =
10$--20 \citep{Lamb_Reichart_00}, and should thus have Population III progenitors. These
afterglows will be both highly time-dilated and severely reddened. They provide an
important tool to probe the evolution of the ionisation state and metallicity of the
Universe, and the physical properties of the first stars to form. Additionally, evidence
suggests that some long duration GRBs are associated with Type Ib/Ic supernovae
\citep[e.g.,][]{Woosley_Bloom_06}.

Even in the current Swift era, there is a paucity of uniform optical afterglow
observations of GRBs. Many of the interesting gamma-ray triggers from Swift and other
satellites, including high-redshift candidates, are not adequately observed. Also, there
is great interest in detecting orphan afterglows that are expected to occur at higher
rates than normal afterglows \citep{Totani_Panaitescu_02,Zou_e_07} and provide valuable
information on beaming.

\subsubsection{Other Facilities}
The Swift gamma-ray satellite, which currently finds \si100~GRBs per year and provides a
location accuracy of \si3$''$ within a few minutes for a large fraction of these
\citep{Gehrels_e_04}, has an orbital lifetime to \si2020 and is funded through 2012 with
extensions possible after that. The Gamma-ray Large Area Space Telescope (GLAST) was
launched in January 2008 with a lifetime of 5--10~years and is expected to find
\si50~GRBs per year with a location accuracy of \si10$'$ with the LAT instrument, and a
further \si150 GRBs per year with an accuracy of \si10\degs\ with the GBM instrument
\citep{Von_Kienlin_e_04}. Future GRB space missions, such as the Space multi-band
Variable Object Monitor (SVOM), the Explorer of Diffuse emission and Gamma-ray burst
Explosions (EDGE), and the Energetic X-ray Imaging Survey Telescope (EXIST) are each
expected to find several hundred GRBs per year with location accuracies ranging from
1--10$'$ \citep{Den_Herder_e_07,Band_e_08}.

\begin{figure}[t]
\begin{center}
\includegraphics[width=7.5cm]{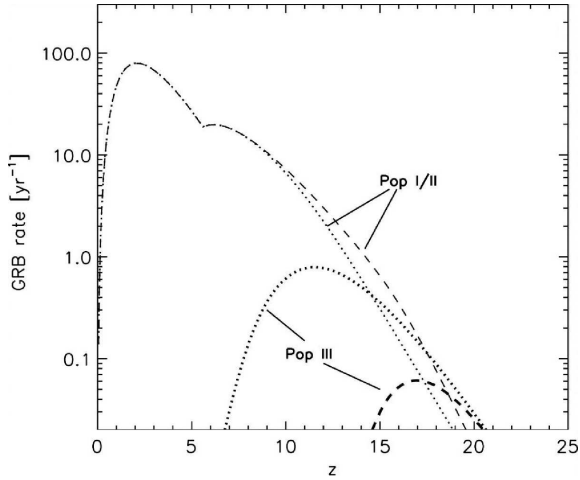}
\caption{Predicted GRB rates observed by Swift as a function of redshift for
Population I, II, and III progenitors. The Population III rate includes strong
and week feedback models, dashed and dotted lines respectively. From \citet{Bromm_Loeb_06}.}\label{GRB_rate}
\end{center}
\end{figure}

\citet{Salvaterra_e_08} have recently estimated the expected number of high-redshift
satellite GRB detections based on models of the luminosity function and formation rate
for long duration GRBs. These models predict that Swift should find 3--16~GRBs per year
at $z\geq6$, and 0.3--0.9~GRBs per year at $z\geq10$. Models from \citet{Bromm_Loeb_06},
shown in Figure~\ref{GRB_rate}, are more optimistic, and also predict that the number of
high-redshift GRBs arising from Population III progenitors could be significant. The
analysis of \citet{Salvaterra_e_08} showed that GLAST is only expected to find a few GRB
per year at $z\geq6$, and has a low probability of finding any GRBs at $z\geq10$. SVOM,
EDGE, and EXIST are more appropriate to high-redshift detections, however, and should
find 2--50 per year at $z\geq6$, and 0.1--3 per year at $z\geq10$.

There are a number of ground-based facilities that are capable of responding to such GRB
satellite alerts and performing optical/near-infrared follow-up. These include dedicated
telescope networks, such as ROTSE, dedicated survey telescopes, such as VISTA, and
general purpose 8--10~m class telescopes (Keck, Gemini, VLT, etc). The applicability of
such facilities to high-redshift GRB afterglow detection is limited, however, thermal infrared sensitivities (due to atmospheric thermal emission), and the availability and wavelength coverage of near- and mid-infrared instruments on 8~m class telescopes. This is evidenced by that fact that although Swift has probably detected many high-redshift GRBs according to the predictions of \citet{Salvaterra_e_08}, only \si30\% of Swift bursts have redshift determinations. Many bursts with stretched out gamma-ray light curves that appear to be at high redshift have either not been optically followed-up or have been followed up only at visible wavelengths with no detection. There is thus a great potential here that needs rapid infrared follow-up to realize.

There are also a number of ground-based projects with aims to independently search for
GRB orphan afterglows, such as LSST and VISTA. With very large fields-of-view, these
facilities are well suited to searching for such objects. Again, however, long wavelength cut-offs and thermal infrared sensitivities will preclude these facilities from finding
orphan afterglows originating from very high-redshift sources.

\begin{figure}[t]
\begin{center}
\includegraphics[width=7.5cm]{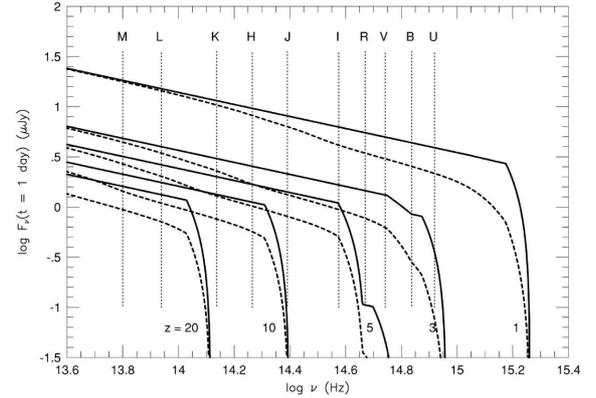}
\caption{Spectral flux distribution of GRB970228, as observed one day after the burst,
after transforming it to various redshifts and extincting with a model of the
Ly-$\alpha$ forest. From \citet{Lamb_Reichart_00}.}\label{GRB_flux1}
\end{center}
\end{figure}

PILOT will not be competitive with LSST for finding low-redshift orphan-afterglows in the visible, as spatial resolution is not critical and LSST is \si25~times faster to a given
depth. PILOT may, however, be uniquely suited for orphan afterglow searches in the
near-infrared: it should be \si8~times faster than VISTA at \kb\ and has a wider
wavelength range. These objects are expected to be very rare, however. More experimental
data are needed to constrain the expected rates for orphan afterglows; none have so far
been detected, and there are large uncertainties (3--4~orders of magnitude) in the
predictions on the expected number density
\citep[e.g.,][]{Totani_Panaitescu_02,Zou_e_07}.

\subsubsection{Observing Strategies}

The key role for PILOT will be to respond quickly to either satellite or ground-based GRB detections, and follow up rapidly in the near- to mid-infrared (from 1--5~\mm). The high
sensitivity in the thermal infrared, the wide infrared wavelength range, and the
potential to respond quickly and observe for long periods, means that PILOT is well
suited to finding high-redshift GRBs. Space-based infrared telescopes, such as JWST, have an enormous advantage in terms of infrared sensitivity, but have a slow response time,
and their smaller field-of-view can make burst location problematic.

\begin{figure}[t!]
\begin{center}
\includegraphics[width=7.5cm]{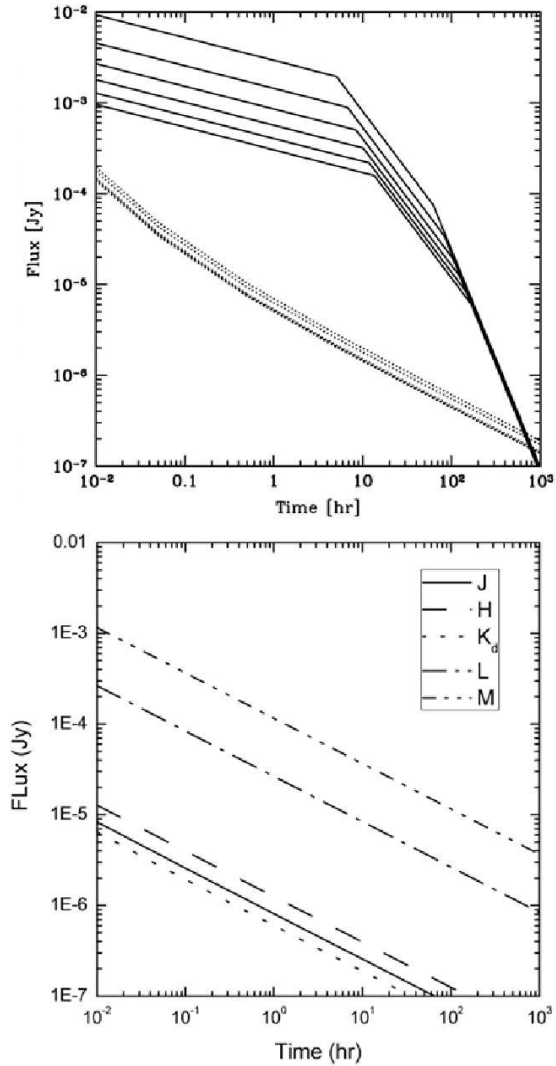}
\caption{Top: high-redshift GRB afterglow flux (solid curves) as a function of time since
the GRB explosion. Dotted curves show the $R = 5000$ spectral detection threshold for the
near-infrared spectrometer on JWST. Both sets of curves show redshift $z = 5, 7, 9,
11, 13, 15$ (top to bottom respectively). From \citet{Barkana_Loeb_04}. Bottom: PILOT
PNIRC detection threshold in $J, H, \kd, L, M$ bands. For both plots, the integration
time is 20\% of the time since the burst.}\label{GRB_flux2}
\end{center}
\end{figure}

While the effects of distance and redshift act to reduce the flux of a GRB at a fixed
wavelength, time dilation acts to increase it at a fixed time of observation from onset.
Thus, in a given spectral band, there is little decrease in the intensity with redshift
at a fixed time after the explosion. Based on fitting the spectral energy distribution to one GRB, \citet{Lamb_Reichart_00} estimate that the flux, one day after burst, will range from 10--20~\mj\ in \emph{K}, \emph{L}, and \emph{M} bands for redshifts from $z = 3$--20 (except for \kb\ at $z = 20$, when the flux is shifted out of the band), as shown in
Figure~\ref{GRB_flux1}. In 5~minutes of observation, PILOT should be able to detect these objects with a $\emph{SNR} > 10$ at \kd\ and a $\emph{SNR}\approx1$ at \emph{L} and
\emph{M}. At earlier times, with the flux estimated to vary as $t^{-4/3}$, higher
\emph{SNR}s would be anticipated. Within the first hour of the burst they should readily
be detectable. This suggests a filter sequence for a search for GRBs, from \emph{J},
through \emph{H}, \kd\, \emph{L}, to \emph{M}, starting with 1--2~minutes at \emph{J} and \emph{H}. If no object is detected, then 5--10~minutes would be spent at \kd, rising to
an hour at \emph{M}. A GRB found only in \emph{M}-band would have $z > 30$. Later models
from \citet{Barkana_Loeb_04} shown in Figure~\ref{GRB_flux2}, and also \citet{Gou_e_04}
and \citet{Inoue_e_07}, give a significantly higher expected flux for high-redshift GRBs; events at this level could thus be followed for several days.

As shown here, PILOT should be capable of detecting very high-redshift GRB afterglows.
The likelihood of such detections will ultimately be determined by the number of
high-energy satellite alerts. A few hundred satellite detections per year are expected
over the whole sky. If PILOT could follow-up in the near-infrared every GRB alert in the
observable sky at Dome~C during winter (i.e., about one a week) within \si10~minutes, it
should find several $z > 6$~GRBs per season and at least one $z > 10$~GRB in a few years.
The distance to these objects can be determined via the Lyman~$\alpha$ drop-out
wavelength. This will provide a tight constraint on the occurrence of GRBs in the early
Universe. Additionally, as a by-product of other PILOT wide-field \kd\ survey projects,
the rate of occurrence of high-redshift GRB orphan-afterglows will be probed to lower
number densities than possible with any other facility.

%%-----------------------------------------------------
%%%%%%%%%%%%%%%%%%%%%%%%%%%%%%%%%%%%%%%%%%%%%%%%%%%%%%
\section{Assembly of Structure}
\subsection{High-Redshift \kd\ Galaxy Surveys}
\subsubsection{Impact}
The study of galaxy formation and evolution through cosmological history was
revolutionised in 1995 with the observation of the first of the Hubble Deep Fields
\citep[HDF;][]{Williams_e_96}. Staring at a field in Ursa Major about 2$'$ across, over a period of ten days and through four broadband filters ranging from the near-ultraviolet
to the near-infrared, over 3000 galaxies were identified. While many were clearly
recognisable as spirals, there were also a larger proportion of disturbed and irregular
galaxies than seen in the local Universe. The HDF also showed that the star formation
rate was clearly higher in the past, and that galaxy collisions and mergers were more
common than today. Subsequent studies have now made it clear that galaxies cover the sky
with number densities of up to one per square arc-second, star formation rates peaked
around $z = 1$--4, and our view of the Universe at this epoch is one of great
disturbance.

Such deep fields are, however, inherently biased. In the optical they sample the
rest-frame ultraviolet-light for galaxies that have $z > 1$. This makes them sensitive to where the most active star formation is taking place at these epochs; i.e., the
ultraviolet radiation produced by young, luminous OB stars spread about their host
galaxies (with an unknown distribution) dominates the view we see. The HDF at high
redshift is thus dominated by Lyman-break galaxies (LBG).  The light from the bulk of the stellar population---older, redder stars---is not seen, for this is redshifted out of the optical bands.  It is necessary to look into the infrared to see this light.

The longest waveband where deep observations can be made from ground-based observatories
is \kb\ (i.e., from 2.0--2.4~\mm); at longer wavelengths the thermal emission from both
the atmosphere and the telescope rises rapidly. In order to see the bulk of the light
from evolved stars, it is necessary to observe at wavelengths beyond the rest-frame for
the 4000~\AA\ and Balmer breaks. For $z > 3$, this means \kb\ is essential.

The HST-NICMOS observations of the Hubble Deep Field South (HDF-S) identified 6\% of the
galaxies as having a photometric redshift $z > 5$, with another 3\% as being evolved E/SO galaxies. Most of these galaxies could have been identified from purely optical deep
fields. The deepest \kb\ image to date was taken in the HDF-S field, as part of the
VLT-FIRES survey \citep{Labbe_e_03}. As the authors comment, there is a striking variety
in optical-to-infrared colours across the field, especially for the fainter (more
distant) objects. Many of the sources with red colours have photometric redshifts $z > 2$ and are candidates for being massive, evolved galaxies. They would not have been
identified by the traditional \emph{U}-band dropout technique applied to the optical
HDFs, for they are too faint in these bands.

\begin{table*}[t!]
\begin{center}
\caption{Comparison of survey capabilities in \kb.}\label{Tab_gal}
\begin{tabular}{p{1.8 cm} c c c c c c c}
\\
\hline
Telescope & Diam & FWHM	    & Pixel Size &	Sensitivity & Survey Area	  &	No. gal.       &	Mass Limit \\
          & (m)	 & (arcsec) & (arcsec)   &	(\mab)      & (arcmin$^{2}$)  &	arcmin$^{-2}$  & (\msun) \\
\hline
PILOT$^a$ &	2.5	 & 0.3      &	0.15     &	26.8        & 5000            &	190            &   3 $\times 10^{9}$\\
UKIRT$^b$ & 3.8	 & 0.8      &	0.40     &	25.0        & 2800            &	65	           &   2 $\times 10^{10}$\\
Blanco$^c$&	4.0	 & 1.2      &	0.31     &	23.0        & 400             &	20	           &   1 $\times 10^{11}$\\
VLT$^d$	  & 8.2  & 0.5      &	0.15     &	25.7        & 6               &	100	           &   8 $\times 10^{9}$\\
HST$^e$	  & 2.4  & 0.2      &	0.08     &	23.1        & 1               &  20	           &   1 $\times 10^{11}$\\
\hline
\end{tabular}
\medskip\\
\end{center}
$^a$ For a PILOT deep field survey with the PNIRC camera using 15 hour integrations per
field, spread over 1000 hours of total observing time, with 70\% observing efficiency.\\
$^b$ For the planned Ultra Deep Survey (UDS) using \si300 nights observing on the UKIRT
telescope as part of the UKIDSS Survey \citep{Lawrence_e_07}.\\
$^c$ From the Multi-wavelength Survey by Yale-Chile (MUSYC) on the CTIO Blanco telescope \citep{Quadri_e_07}.\\
$^d$ From the Faint Infrared Extragalactic Survey (FIRES) on the VLT telescope \citep{Labbe_e_03}.\\
$^e$ From the Hubble Deep Field South survey using the near-infrared camera and multi-object spectrograph (NICMOS) on the HST \citep{Yahata_e_00}.\\
\end{table*}

While it appears that around two-thirds of the star formation in the Universe has
occurred since $z<2$, the remaining third has occurred at earlier epochs
\citep[e.g.,][]{Glazebrook_e_04}. Much research in this field is directed to
understanding when and how these stars formed, and what this implies for hierarchical
models of galaxy formation. It is important to ascertain which population of galaxies
hosts the star formation as a function of cosmic epoch. PILOT can address this directly
by measuring stellar masses at high redshift. To study the evolved galaxy population at
$z>3$ in order to address these questions requires deeper measurements in the \kb, over
wider fields-of-view, to both better sample the population of old galaxies to lower mass, and to overcome cosmic variance in the fields. At temperate-latitude observatories the
rising thermal background limits the achievable sensitivity, requiring a much larger
aperture. However, from the Antarctic plateau, the extreme cold dramatically lowers the
thermal background, leading to greatly improved sensitivities for the same-sized
telescope.

\subsubsection{PILOT Deep Fields}
\begin{figure*}[t!]
\begin{center}
\includegraphics[width=7.5cm]{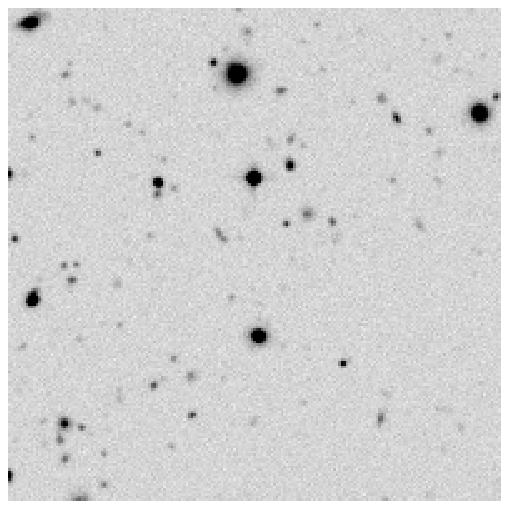}
\hspace{0.5cm}
\includegraphics[width=7.5cm]{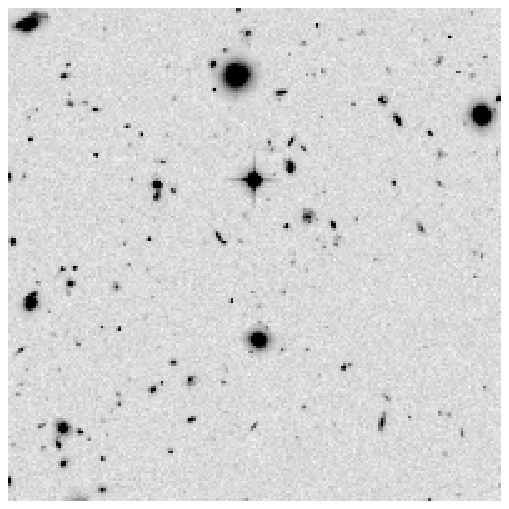}
\caption{Left: Simulated ultra-deep \kb\ image, calculated using BUCS, as could be
obtained by the UKIRT-UDS, achieving a 5~$\sigma$ galaxy detection limit of $\mab =
25.0$. A $30'\times30'$ field is shown, with spatial resolution 0.8$''$.  Right: A
similar simulation for a \kd\ image, as could be obtained by PILOT in 15~hours of
integration, achieving a 5~$\sigma$ galaxy detection limit of $\mab = 26.8$. The same
field is shown, with a spatial resolution of $0.3''$. Roughly 200~galaxies per
arcmin$^{2}$, three times as many as the UKIRT-UDS, are detected. The complete PILOT
survey could cover \si5000~arcmin$^{2}$ at this depth. }\label{gal_sim}
\end{center}
\end{figure*}

We now consider the performance possible for a \kd\ deep field obtained from Antarctica,
compared with deep fields possible from mid-latitude locations. We have applied the
Bouwens Universe Construction Set \citep[BUCS;][]{Bouwens_e_05} software package in order
to generate simulated deep galaxy fields. BUCS makes use of real galaxy templates
extracted from deep multi-colour HST observations, including the GOODS (Great
Observatories Origins Deep Survey) field. It places these templates on the simulated
frames with the same surface densities as found in the input samples. In creating the
output images, shown in Figure~\ref{gal_sim}, BUCS recalculates the appearance of each
galaxy using the best-fit pixel-by-pixel spectral energy distributions, and then
resamples this onto the output frame. The results are then smoothed to the
point-spread-function (PSF) of the output image, and noise is added to reproduce the
specified background noise level. By preserving both the surface densities and the
pixel-by-pixel morphological details of each input object, BUCS is able to produce
model-independent representations of each simulated field.

We compare the parameters of several current surveys in Table~\ref{Tab_gal}. The number
density of galaxies to a given magnitude limit, as predicted by the models, is also
listed in Table~\ref{Tab_gal}. PILOT could, for instance, go 1~magnitude fainter than the
VLT-FIRES survey over a 3~orders of magnitude larger area, or 2~magnitudes deeper than
the UKIRT-UDS over twice the area, detecting an order of magnitude more galaxies.  In
both cases the spatial resolution would be 2--3~times better.

The mass limits reached by a particular survey are also listed in Table~\ref{Tab_gal}.
While obviously this limit depends on many assumptions about the stellar light of a
galaxy and its evolutionary state, we make use of the work of \citet{Van_Dokkum_e_06} to
list an indicative mass limit for which a particular survey will be complete, in the
redshift range $2 < z < 3$. For $z < 1$ the limiting mass will be less than 10 times
smaller, whereas for $z > 4$ it would be about 3 times larger. To provide a comparison
between surveys, MUSYC reached to $10^{11}$~\msun\ masses, while FIRES reached an order
of magnitude lower in mass, but with an order of magnitude smaller field size. PILOT
could reach to \si$3\times10^{9}$~\msun\ over a 1--3~orders of magnitude larger field
size than these other surveys.

An alternative performance metric is to consider how large an area might be surveyed to a ``deep" limit, as opposed to an ``ultradeep" limit, in comparison to the wide-field
UKIRT-DXS (Deep eXtragalactic Survey). This latter survey with the UKIRT will cover
35~\degsq\ of sky, to a 5~$\sigma$ detection limit of $\mab = 23.0$~mags (2~mags
shallower than the UKIRT-UDS). It is slated to take 118~nights of telescope time, in
comparison to 296~nights for the UDS. With PILOT, integrating for 1~hour per field, the
5~$\sigma$ detection limit would be $\mab = 25.3$~mags, and 20~\degsq\ could be covered
in \si100~nights (assuming, as does UKIRT, 10~hours of observing per night with an
efficiency of 70\%). As well as going 2.3~magnitudes deeper than the UKIRT-DXS over a
comparable area, such a PILOT-DXS would yield an image resolution of 0.3$''$ compared to
0.8$''$. Such wide areal coverage is also well beyond what could be covered by any survey conducted using JWST, although, of course, JWST will go much deeper ($\mab \approx30$ at
\kb) than any proposed ground-based telescope could ever achieve.

\subsubsection{The First Stellar Populations}
The discovery of galaxies at redshift $z > 6$ with well-developed Balmer breaks
\citep{Eyles_e_05} is a challenge to galaxy formation theories, since these breaks take
more than 200~Myr to form, and the Universe was only 1000 Myr old at that time. In any
case, these must be the galaxies whose light reionised the Universe at $z > 10$. There is an obvious case to (a) determine the space density of these galaxies at $z = 6$; (b)
measure the decline rate in star forming galaxies over the range $3 < z < 6$; and (c)
find the very earliest stellar populations in the Universe. The galaxies found by
\citet{Eyles_e_05} were only detected because they also have Lyman breaks (and hence
ongoing star formation); it is a puzzle as to why such a large fraction (40\%) of these
Lyman break galaxies also have Balmer breaks. In the nearby Universe, galaxies exist with recently formed Balmer breaks but no current star formation (E+A galaxies), so we need a
method that can find all Balmer break galaxies at these redshifts without depending on
star formation.

\begin{figure*}[t!]
\begin{center}
\includegraphics[width=7cm]{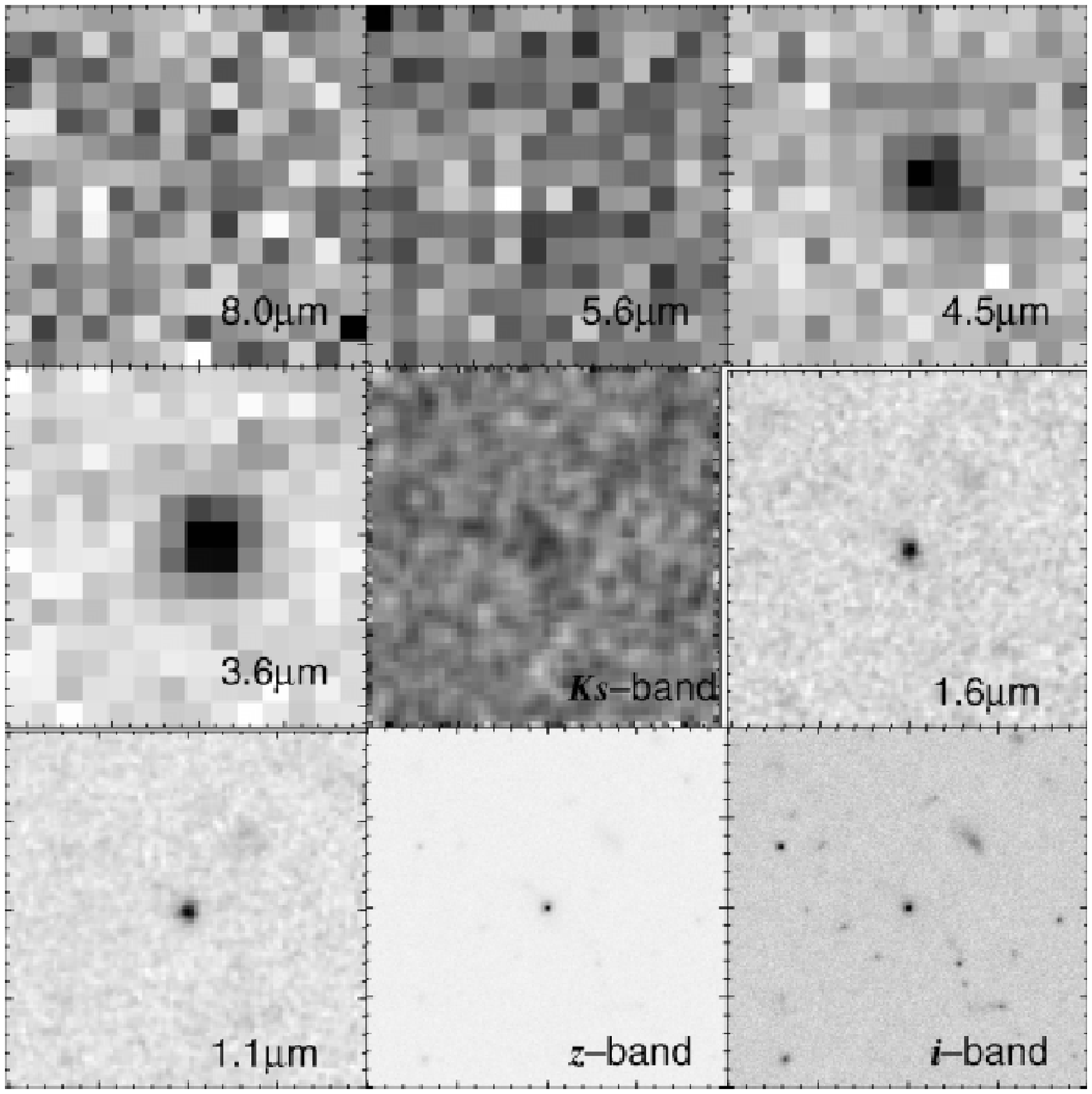}
\includegraphics[width=7.5cm]{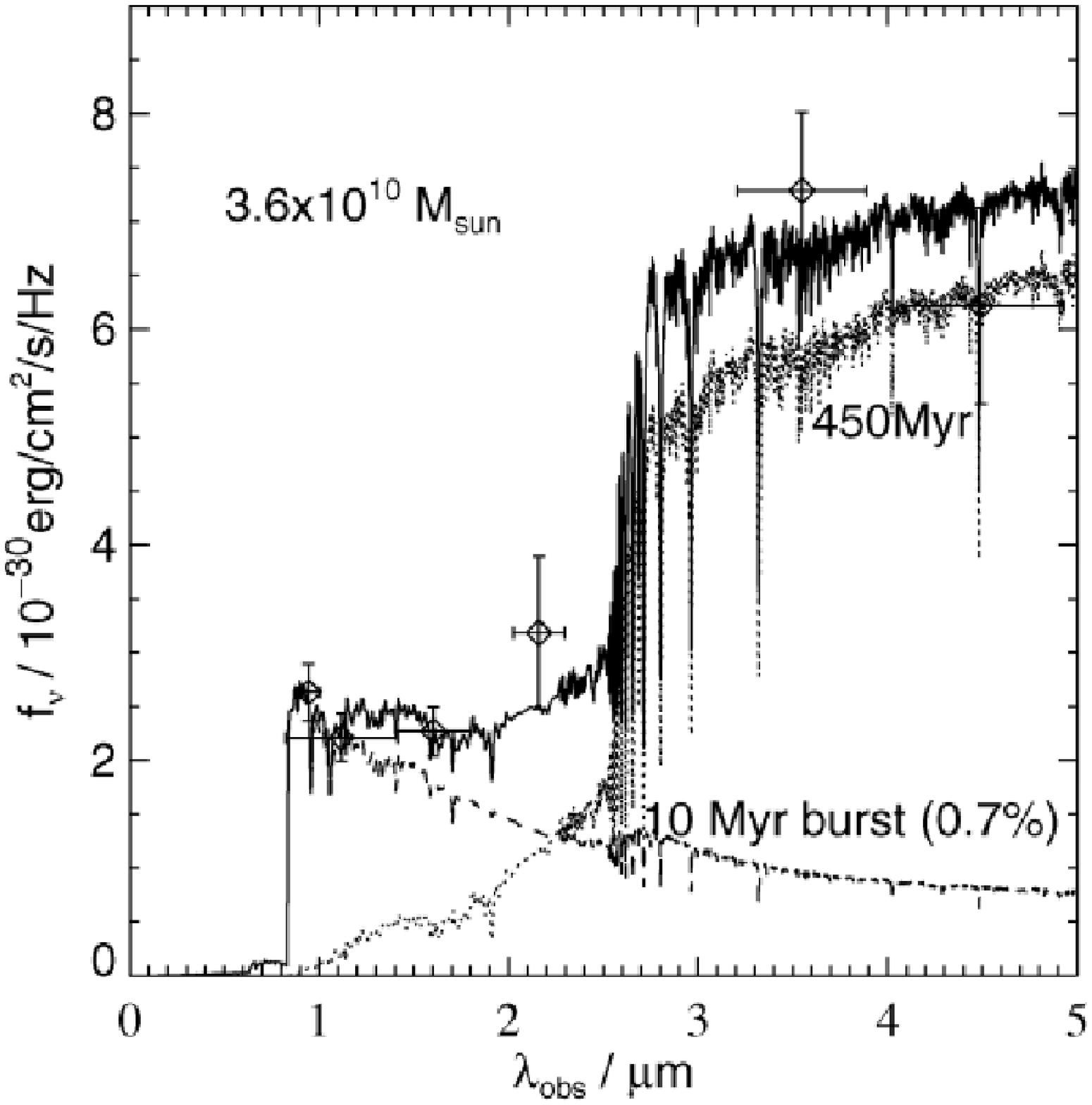}
\caption{Left: postage-stamp images (each 8$''$ across) at different wavebands of a
spectroscopically confirmed $z = 5.83$ Balmer break galaxy. Data comes from HST-ACS
(\emph{i}- and \emph{z}-band), HST-NICMOS (\emph{J}- and \emph{H}-band), VLT-ISAAC
(\emph{K}-band) and Spitzer-IRAC (3.6--8~\mm). Right: best fit synthetic spectrum,
combining star forming and 450~Myr populations. From
\citet{Eyles_e_05}.}\label{gal_break}
\end{center}
\end{figure*}

Figure~\ref{gal_break} shows the data for one of the \citet{Eyles_e_05} galaxies. The
Spitzer data at 4.5 and 3.6~\mm\ are very good. The data at 1.6~\mm\ and shorter, from
HST, are also good. The glaring problem is the \kb\ data that are essential for
determining the break amplitude and hence the age and metallicity; these data have poor
\emph{SNR} despite representing 4--5~hours integration, in the best conditions, on an 8~m telescope. While the large breaks have been attributed to Balmer breaks, it is possible that emission lines in the Spitzer bands could also be responsible.

The Spitzer warm mission is now funded for at least two years; the Infrared Array Camera
(IRAC) will continue to work as before in its two shortest bands. It happens that, for
this project, the IRAC sensitivities are almost perfectly matched to those of PILOT at
\kd. IRAC reaches 0.5~\mj\ in 1~hour at 3.6~\mm\ and 0.5~\mj\ in 4~hours at 4.5~\mm. The
expected [$\kd - 3.6$~\mm] flux ratio is a factor 3--4 \citep[e.g.,][]{Van_Dokkum_e_07}.
PILOT+PNIRC (with a field 5 times larger) will take 4~hours to reach the required
0.15~\mj\ at \kd. PILOT can thus provide matching \kd\ photometry (and accurate
positions) for essentially all IRAC detections, in similar timescales. The spectral
signal is very distinct---a 1.5 mag jump and then flat.

The redshift range over which we could detect galaxies by this method corresponds to when the Balmer break is between the \kd\ and 3.6~\mm\ bands, i.e., $z = 5$--7. Note that
\citet{Mobasher_e_05} claim sources at $z\approx6.5$, but this is disputed
\citep{Munoz_Loeb_08}. PILOT also provides a window for detecting galaxies with the break between the 3.6~\mm\ and 4.5~\mm\ bands (as 3.6~\mm\ dropouts) at a redshift $z =
9$--11.5. Such galaxies are not thought to exist, but this needs to be demonstrated.

\citet{Eyles_e_05} detected 6~galaxies with $z\approx6$ in an area of 150~arcmin$^{2}$.
The number of galaxies that are undetected because they have no star formation is
unknown. To determine a luminosity function, and understand the evolution of the
population, requires several hundred to be found at least, necessitating a survey of a
few square degrees. Such a survey would take \si500~hours with PILOT.

For PILOT, the science payoff is obvious. All that is needed is for Spitzer to undertake
some of its major extragalactic surveys far enough south to be observable by PILOT. For
Spitzer, which will probably be decommissioned before PILOT is built, the payoff is less
immediate. However, the existence of BLANCO Dark Energy Survey (DES) \emph{g}, \emph{r},
\emph{i}, \emph{z} photometry, and the proximity of the South Ecliptic Pole (which is
just outside the DES area) provides an argument to choose a suitable area now.

%%-----------------------------------------------------
\subsection{Witnessing Galaxy Cluster Assembly at Moderate Redshift}

\begin{figure*}[t!]
\begin{center}
\includegraphics[width=15cm]{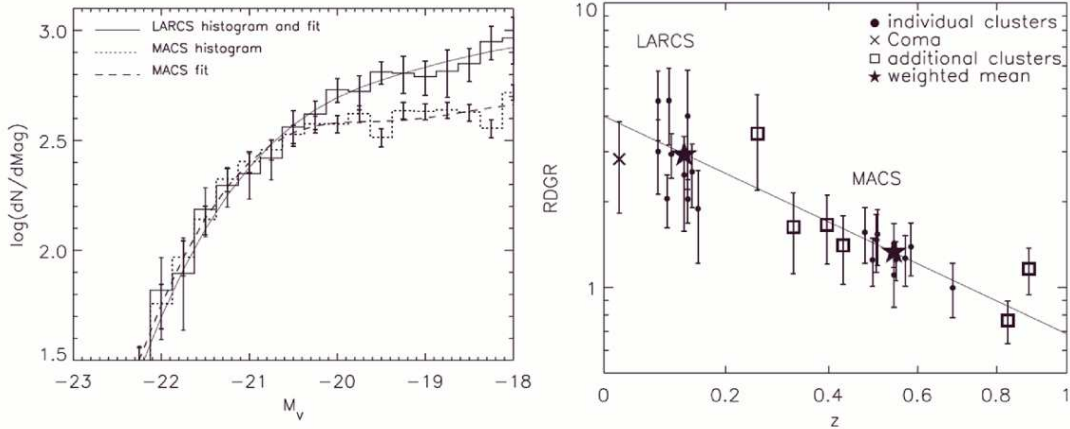}
\caption{Left: luminosity functions for red-sequence galaxies in low (LARCS) and high
(MACS) redshift samples demonstrating faint red galaxy excess. Right: variation in
Red-sequence Dwarf to Giant Ratio (RDGR) in clusters versus redshift. From
\citet{Stott_e_07}.}\label{cluster}
\end{center}
\end{figure*}

\subsubsection{Impact}
The field of galaxy cluster studies has entered a renaissance with modern data sets such
as 2dFGRS and SDSS being made publicly available. There are thousands of clusters now
catalogued, each with tens (if not hundreds) of associated galaxies with photometric and
spectroscopic data readily available for download. These datasets have allowed us to gain
an unprecedented picture of how galaxy clusters grow through the accretion of smaller
groups of galaxies as they fall into the cluster potential anisotropically along
filaments.  Indeed, to pick just one example from many, we now know that there is a
critical density (or clustercentric radius) for galaxy formation
\citep[e.g.,][]{Gomez_e_03,Lewis_e_02}. However, it is worth noting that the median
redshift for the SDSS and 2dFGRS surveys is still low: in the region of $0.10 < z <
0.15$. The relatively smaller amount of data that we have on significantly
higher-redshift clusters comes usually from only the most massive clusters (since they
are easier to detect, e.g., at X-ray wavelengths) and have typically not been surveyed in
a systematic wide-field manner.

There are therefore two obvious gaps in our dataset: the existence of a long redshift
baseline to delineate the effects of cosmic time on the galaxy populations of clusters,
and a range in the mass of galaxy clusters at different epochs to delineate the role of
environment.

Here, we propose a suite of observations to observe high-redshift clusters and their
environment in a homogeneous manner. We will be surveying the field regions and filament
regions that are fuelling the cluster growth.

So called third-generation X-ray surveys such as the XMM distant clusters project have
generated thousands of possible detections of $z > 0.5$ galaxy clusters
\citep[e.g.,][]{Fassbender_08}. Observations of this high-redshift sample would provide a
large redshift baseline and some degree of mass range at high redshift to compare with
our lower-redshift cluster samples \citep[see e.g.,][]{Stott_e_07}. This would yield
important information in a number of areas, including, but not limited to:
\begin{itemize}
\item the evolution and build-up of the colour magnitude relation with redshift,
    including how elliptical and lenticular galaxies form and evolve; this would also
    test the monolithic collapse scenario versus the assumed hierarchical assembly
    model. We should be seeing a large amount of scatter in high-redshift
    colour-magnitude diagrams during the peak epoch of star formation at $z \approx
    1$,
\item the role of AGN feedback over a large epoch in look-back time,
\item the formation epoch and subsequent evolution of brightest cluster (cD class)
    galaxies and how they relate to their local environment,
\item the amount of recent major cluster-cluster merging activity to test predictions
    from modern semi-analytic simulations (e.g., the Millennium run),
\item the morphology-density-star formation-colour relationships at high redshift
    \citep[cf.,][]{Smith_e_05}, and hence the relationship to the Butcher-Oemler
    effect,
\item quantification of the effect of downsizing (i.e., star-formation occurring in
    smaller mass galaxies with decreasing redshift) out to large redshift
    \citep[cf.,][]{Cucciati_e_06} and, as illustrated in Figure~\ref{cluster}, the
    lack of faint red galaxies at this high-redshift epoch
    \citep{Stott_e_07,De_Lucia_e_04}, and
\item the role of the cluster environment within a supercluster context
    \citep{Stott_e_07,Stott_e_08}.
\end{itemize}

\subsubsection{Other Facilities}

Most previous observations of high-redshift clusters have taken place in a limited
context. The number of clusters studied has been small, and hence real local
cluster-to-cluster variations can dominate the ensemble. The field-of-view has been
moderately small (meaning that only the central Mpc or less might be viewed---especially
true of HST where complex tessellation patterns are required to map out clusters to
significantly large radii) thereby denying us a view of the filaments and infalling
groups that will come to accrete on to the cluster potential in the next few Gyr and the
location of a given cluster within a supercluster environment. Finally, the resolution of
existing surveys is poor, such that individual galaxy morphologies may only be crudely
estimated.

PILOT offers a unique blend of a large field-of-view, allowing us to probe to significant
radii and density regimes in an homogeneous manner; an excellent resolution, important
for morphological dissection; and excellent prospects of obtaining a large amount of data
across the southern sky without having to wait for the telescope to re-visit a given
pointing several times in order to build up good photometry.

\subsubsection{Observations} We note from the outset that cluster and supercluster
observations could readily ``piggy-back" on other potential science drivers and could
therefore represent an excellent return on investment.  To perform the kind of science
outlined here, we need to observe sufficiently far down the luminosity function to
effectively probe the fainter galaxies where much of the star formation would be likely
to occur---$M^{*}+2$ would be about ideal in the first instance to determine the
population, or lack thereof, of faint red colour-magnitude ridge-line objects. Ideally,
we require wide-field observations made in a homogeneous manner, preferably in
photometric conditions. At low redshifts, several authors have made progress by actively
pursuing galaxy evolution to very large clustercentric radii
\citep{Wake_e_05,Pimbblet_e_06}---certainly out to \si8~Mpc. At high redshifts (say $0.8
< z < 1.5$), the distance scale would be as much as 8~kpc per arcsec meaning that the
distance on the sky would be much less than a degree for matching to baseline $z \approx
0.1$ observations. An $M^{*}$ galaxy at $z = 0.8$ corresponds roughly to $\mab = 24.5$ at
\vb\ (or $\mab = 25.5$ by $z = 1.0$). According to \citet{Romer_e_01} and
\citet{Fassbender_08} there are about 750~clusters at $z < 0.6$ and perhaps up to 100
more in the range $0.6 < z < 1.5$ identifiable from the XMM cluster survey.

A proposed project for PILOT is to image a sample of 100~galaxy clusters in the range
$0.6 < z < 1.5$. This is a large enough sample to be homogeneous enough to beat down the
local cluster variations. The image size of the PILOT wide field visible camera, PVISC,
is large enough that the majority of clusters can be observed with a single pointing.
Three bands in the visible (\emph{r}, \emph{i}, and \emph{z}) would be chosen to straddle
the 4000~\AA\ break at these redshifts, and to give a comfortable dynamic range to derive
colours. The expected sensitivities of PVISC implies \si1~hour of observation per filter
per cluster. The complete survey would thus take \si300~hours. Ideally, follow-up
observations of the same clusters in the near-infrared \emph{J}, \emph{H}, or
\emph{K}-bands with the PNIRC camera would give an even larger dynamic baseline. For such
a small investment in observing time, this project should represent the largest sample of
X-ray selected clusters examined at these redshifts.

%%%%%%%%%%%%%%%%%%%%%%%%%%%%%%%%%%%%%%%%%%%%%%%
\section{Dark Matter and Dark Energy}
%%-----------------------------------------------------
\subsection{Weak Lensing}
\subsubsection{Overview}

Cosmic shear offers a relatively clean way to measure the evolution of the power
spectrum, and hence the equation of state, of the Universe. The observations are in
principle simple: since large-scale structure has a quadrupolar lensing effect on
background galaxies, we can map this structure by measuring the shapes of faint galaxies.
Large numbers (\si10$^{9}$) are needed, to extract the faint lensing signal from the much
larger intrinsic shape variation, and to overcome cosmic variance. If we know the
approximate distances of these faint galaxies, and can measure the lensing for separate
distance classes, then we can measure the evolution of clustering, which in turn depends
directly on the equation of state. The measurement is clean because we measure mass
fluctuations in the linear regime. Weak lensing has been recognised as the most promising
route for determining the characteristics of dark energy by NASA, NSF and ESA.

Because of its importance, a large number of weak-lensing surveys are currently planned.
From the ground there is: the Dark Energy Survey (2009), KIDS on VST (2009),
HyperSuprimeCam on Subaru (\si2010), PanSTARRS-4 (\si2012), and LSST (\si2014). From
space there is DUNE (now part of EUCLID; \si2018), and JDEM/SNAP (\si2018). For PILOT to
make an impact, it must be better than any of the ground-based surveys, and faster (or
better) than the space missions.

\begin{figure}[t]
\begin{center}
\includegraphics[width=7.5cm]{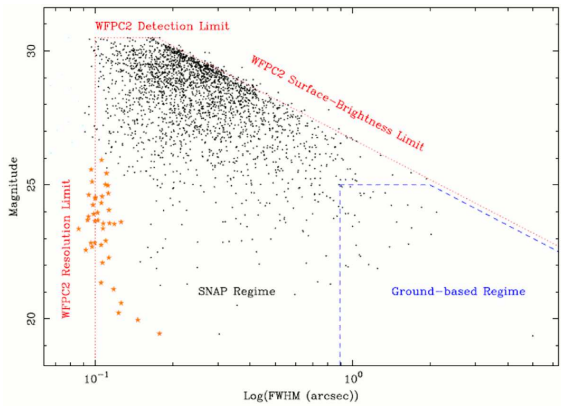}
\caption{Image size for galaxies from the HDF, as a function of AB magnitudes. From
\citet{Curtis_e_00}.}\label{WL_HDF}
\end{center}
\end{figure}

\begin{figure}[h]
\begin{center}
\includegraphics[width=7.5cm]{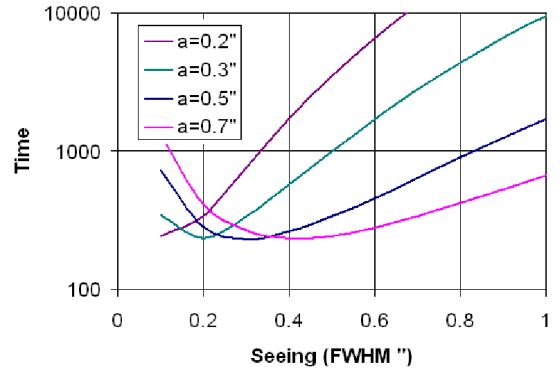}
\caption{Overall weak-lensing survey speed as a function of seeing for galaxies of size
0.2, 0.3, 0.5, and 0.7$''$, assuming that the detector pixels are a fixed fraction of the
seeing.}\label{WL_speed}
\end{center}
\end{figure}

\begin{table*}[t]
\begin{center}
\caption{Parameters for large lensing surveys.}\label{Tab_WL}
\begin{tabular}{p{2.3cm} c c c c c c}
\\
\hline
Telescope/ & Diam    & Giga-	    & Res.$^a$    & Signal$^b$    & Survey  & Year$^d$ \\
Instrument & (m)    & pixels	    & (arcsec)  &     & speed$^c$  &  \\
\hline
MegaCam	      & 3.6	        & 0.34	        & 0.7	       & 0.16	   & 0.11	& 2003 \\
VST	          & 2.6	        & 0.28	        & 0.8	       & 0.12	   & 0.03	& 2009 \\
DES	          & 4	        & 0.5	        & 0.9          & 0.10	   & 0.08	& 2009 \\
PanSTARRS     &	1.8	        & 1.4	        & 0.6	       & 0.20	   & 0.18	& 2009 \\
PanSTARRS-4   &	3.6	        & 1.4	        & 0.6	       & 0.20	   & 0.73	& 2012 \\
\textbf{PILOT}& \textbf{2.5}& \textbf{1.0}	& \textbf{0.2$^e$} & \textbf{0.69}	   & \textbf{3.00}	& \textbf{2013} \\
HyperSupCam	  & 8.2	        & 1	            & 0.7	       & 0.16	   & 1.62	& 2013 \\
LSST	      & 6.5	        & 3	            & 0.7	       & 0.16	   & 3.05	& 2014 \\
DUNE	      & 1.2	        & 0.65	        & 0.23	       & 0.63	   & 3.71	& 2018 \\
SNAP	      & 1.8	        & 0.45	        & 0.15	       & 0.80	   & 9.33	& 2018 \\
\hline
\end{tabular}
\medskip\\
\end{center}
$^a$Resolution is the FWHM image quality determined by the atmospheric seeing conditions,
the telescope correction system, or the diffraction limit. \\
$^b$Signal is the dilution of the lensing signal for a 0.3$''$ FWHM galaxy.\\
$^c$Survey speed is the relatively time to complete a survey of a given area to a given
depth.\\
$^d$Year is the planned date that the survey will begin.\\
$^e$The PILOT project is undertaken in the 50\% best Dome~C seeing conditions.\\
\end{table*}

\subsubsection{Why Antarctica?}
The intrinsic effects of lensing are \si1\% changes in ellipticity. The galaxies at the
required distances and surface densities have $\mab\approx26$ and are smaller than the
best existing ground-based seeing (Figure~\ref{WL_HDF}). Seeing dilutes the lensing
signal by the square of the overall observed image size. This translated directly into
both the signal-to-noise required per galaxy, and the sensitivity to systematic error in
the PSF. It also increases the sky noise, to give an overall integration time per source
varying as the sixth power of the observed image size \citep{Kaiser_e_00}. The overall
survey speed (assuming the pixels are a fixed fraction of the seeing) varies as the
fourth power of the observed image size. For 0.3$''$~FWHM galaxies, this translates to an
order of magnitude gain for Dome~C over temperate sites (as shown in
Figure~\ref{WL_speed}), making PILOT competitive in survey speed with wide-field 8~m
class telescopes such as HyperSuprimeCam or LSST (see Table~\ref{Tab_WL}).

However, an even more serious limitation for temperate sites comes from the PSF
stability. Systematic effects in the variation of the PSF (with time, colour, and field
position) must be understood and modelled at a level better than the diluted lensing
signal. The net effect is that even with 8~m telescopes ``data collected in seeing worse
than 0.8$''$ is of little use for weak lensing" \citep{Kasliwal_e_08}. In practice, the
combined effects of sensitivity and systematics limit the size of galaxies to be
comparable with the seeing, giving a surface density at temperate sites of 10--20
galaxies/arcmin$^{2}$, compared with 50--100 galaxies/arcmin$^{2}$ from space
\citep[e.g.,][]{Kasliwal_e_08}.

At Dome~C, the lensing signal is several times stronger (Table~\ref{Tab_WL}). Moreover,
the better seeing means slower telescope and camera optics, which means excellent image
quality is much more readily achieved and maintained over wide focal planes. It is safe
to assume we can control PSF systematics to the same level, in pixels, as temperate
telescopes. This allows us to reach a surface density \si3~times higher, and a median
depth \si50\% greater. This increases (a) the sensitivity to lensing, (b) the volume
surveyed, and (c) the redshift lever-arm for measuring the evolution of the power
spectrum.

\subsubsection{PILOT Parameters}

We assume a PanSTARRS-type Orthogonal Transfer Array camera with 1.4~Giga-pixels. A
camera with traditional CCD's (e.g., $10\times$~STA1600A detectors, 1.1~Giga-pixels)
would be somewhat less efficient, but would have better PSF constancy (because there are
no anisoplanatic guiding errors). Which of these is the optimal detector technology for a
weak-lensing survey has not yet been determined.

Either way, PILOT would reach $\emph{SNR} = 10$ for $\mab = 25.6$ for 0.3$''$ FWHM
galaxies, with an optimised \emph{r}+\emph{i}+\emph{z} filter in 30 minutes. This gives
\si40 galaxies per arcmin$^{2}$, which allows measurement of the peak of the lensing
power spectrum at $l \approx 5000$--10\,000 (2--4$'$~scales). This scale is where the
greatest sensitivity to cosmological models occurs.

Determining the evolution of the power spectrum requires photometric-redshift
determination. It would not be an efficient use of PILOT time to undertake the required
photometry. However, the Dark Energy Survey \citep[DES;][]{Annis_e_05} will obtain
\emph{g}, \emph{r}, \emph{i}, \emph{z} photometry to limits of $\mab =26.1$, 25.6, 25.8,
25.4, respectively, for 4000~\degsq\ with $\delta < -30$; this is essentially all the
high-latitude sky available to PILOT. This region of sky will be surveyed at millimetre
wavelengths by the South Pole Telescope (SPT). It will also be covered by the VISTA VHS
survey in the infrared, but not deeply enough to be useful. This limits the maximum
redshift to $z \approx 1.2$ (as the Balmer break moves into the $z$-band).

\subsubsection{Proposed Survey}
The proposed project for PILOT is to survey the 4000~\degsq\ DES/SPT area, to a depth
giving 10~$\sigma$ detections of the galaxies at the 5~$\sigma$ DES limit ($\mab\approx
25.6$). Because the speed of such a survey depends so dramatically on image quality, we
propose that this survey only be undertaken in better than median seeing conditions, and
in dark or grey time. The required integration time is about 30~minutes per field, and
the time to cover the entire DES/SPT area is 2500~hours, or a third of the available time
over 4~years.

\begin{figure}[h!]
\begin{center}
\includegraphics[width=7.5cm]{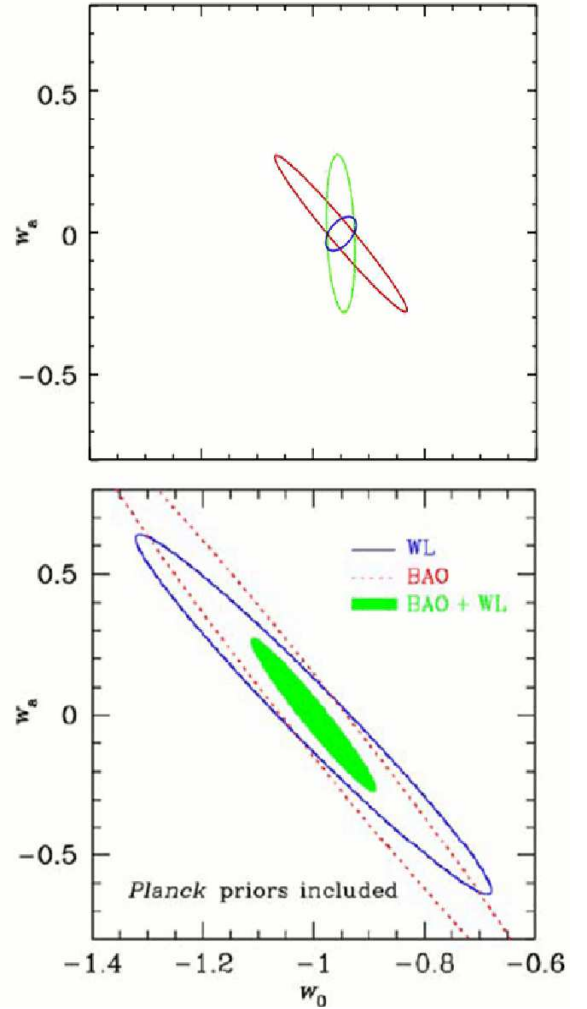}
\caption{Top: constraints on cosmological parameters measured from the DUNE weak-lensing
survey. Different coloured contours correspond to different assumptions for the
photometric redshift errors and for the pivot point $a_n$. From \citet{Refregier_e_06}.
The PILOT errors would be a factor of two larger than the red or green ellipses. Bottom:
constraints from the LSST survey. Contours are constraints from baryon acoustic
oscillations (BAO) and weak lensing (WL). The shaded region combines the two. From
\citet{Tyson_e_06}. In both plots, $\omega$ is parameterised as $\omega(a) = \omega_{n} +
(a_{n} - a)\omega_{a}$.}\label{WL_dark}
\end{center}
\end{figure}

Compared with the proposed DUNE space mission (now part of EUCLID, proposed for 2018), a
PILOT lensing survey looks remarkably good. The overall survey speeds, taking into
account the image quality, sky background, aperture, and field-of-view, are about the
same. PILOT can only access \si25\% of the high Galactic latitude sky, and PILOT will not
be capable of obtaining the deep \emph{Y}, \emph{J}, and \emph{H}-band data proposed for
DUNE. However, this does not affect the main weak-lensing science goals. The constraints
on the equation of state for DUNE are shown in Figure~\ref{WL_dark}. Because the PILOT
survey has a smaller area, the error ellipses are a factor 2 larger. They are still much
better than those calculated for, e.g., LSST \citep{Tyson_e_06}.

%%-----------------------------------------------------
\subsection{Dusty Supernovae}
\subsubsection{Impact}

Supernovae (SNe) are at the intersection of cosmology, galaxy evolution, and stellar
evolution. SNe are studied as the end product of the evolution of massive stars via core
collapse. They eject most of their external layers into the interstellar medium
\citep[e.g.,][]{Arnett_e_89}, and they might be the major dust factory in the Universe
through condensation of dust grains in their ejecta \citep[e.g.,][]{Nozawa_e_03}. Dust
formation in SNe may play a major role in the evolution of galaxies since the dust
absorbs and scatters ultraviolet photons in the far-infrared. Finally, SNe (Type Ia) are
standard candles via which the cosmic acceleration of the Universe can be probed
\citep[e.g.,][]{Schmidt_e_98,Perlmutter_e_99}. This last point is the main driver for
PILOT supernovae studies.

The use of SNe is strongly constrained by the environment in which they explode. If these
very violent events appear in a dusty environment, there are likely to be uncertainties
introduced in the estimation of their distance and absolute brightness from their optical
light curve. This necessarily leads to errors in the derived Hubble diagram. As discussed
by \citet{Burgarella_e_08}, the dust attenuation in magnitudes at \emph{K} is only
one-ninth of that at \emph{V} \citep{Rieke_Lebofsky_85}. This is illustrated by recent
results showing that some supernovae (e.g., SN~2002cv) that are not detected in optical
surveys are easily detected in the infrared \citep{Di_Paola_e_02}. The \vb\ non-detection
suggests that there is a large variation in the dust attenuation of observed supernovae.
This is supported by \citet{Maiolino_e_02} who estimate that the \vb\ dust attenuation of
SN2001db was \si5 magnitudes higher than the average dust attenuation for a more usual
sample of supernovae. \citet{Astier_e_06} stated that ``there is no consensus on how to
correct for host galaxy extinction affecting high-redshift SNe~Ia".

\begin{figure}[h]
\begin{center}
\includegraphics[width=7.5cm]{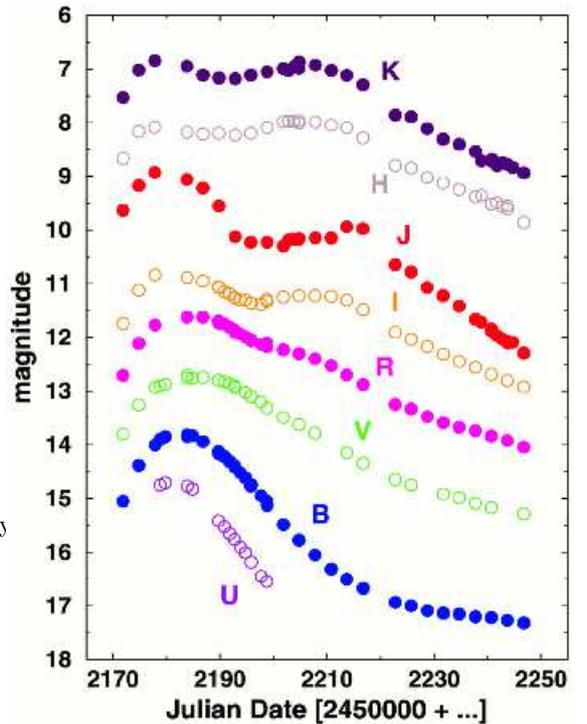}
\caption{Optical and infrared light curves of SN 2001el. Bands have been offset for
plotting purposes. From \citet{Krisciunas_e_03}. These data show that the peak
brightness occurs earlier in the infrared bands than in the visible, but that the
infrared light curves are flatter for a longer period after reaching
maximum.}\label{SN_curve}
\end{center}
\end{figure}

Observing in the infrared should allow more accurate interpretation of SN light curves by
negating dust effects, and hence obtaining more accurate constraints on the cosmological
parameters derived from SN~Ia distance relationships. Recent results have confirmed that
for nearby supernovae, with redshifts up to $z = 0.1$, near-infrared light curves are
excellent standard candles, even without correction for reddening
\citep{Wood-Vasey_e_07,Krisciunas_e_04a}. It has been suggested that specific wavelengths
and redshifts occur that allow a much higher precision distance indicator due to small
K-corrections; for example, where the \jb\ rest-frame is shifted to \kb\ at $z = 0.73$
\citep{Krisciunas_e_04b}. An additional advantage of the infrared for supernovae searches
is that, as shown in Figure~\ref{SN_curve}, the light curves are flatter for a longer
time than in the visible \citep{Krisciunas_e_03,Di_Carlo_04}. This means there is a
longer opportunity to detect these supernovae than in the optical and that the
signal-to-noise ratio will be higher for a longer period. Additionally, with a high
angular resolution it should be possible, with PILOT, to detect supernovae in the nuclear
regions of galaxies where source crowding makes detection with Spitzer, for example,
difficult \citep{Chary_e_05}.

While cosmological studies from SN~Ia provide the prime motivation for PILOT supernovae
searches, the study of supernovae environments would also be enabled by such a search.
The study of dust and dust-related phenomena are important for our understanding of star
formation in the Universe, and in evaluating the star formation rate of SN host galaxies
(for core-collapse SNe). We know that most of the star formation in the Universe at
$z\approx1$ is hidden in dust \citep[e.g.,][]{Takeuchi_e_05}; even ultraviolet selected
galaxies such as Lyman Break Galaxies can present a large dust emission and appear as
Luminous Infrared Galaxies with $10^{11} < L_{dust}/L_{\odot} < 10^{12}$
\citep{Burgarella_e_07}. The detection of supernovae in dusty environments is therefore a
very promising way to explore the relationship between SN processes and the environment
in which they occur \citep[e.g.,][]{Chary_e_05,Pozzo_e_06,Elias-Rosa_e_06}.

\subsubsection{PILOT Observations}

\begin{figure}[h]
\begin{center}
\includegraphics[width=7.5cm]{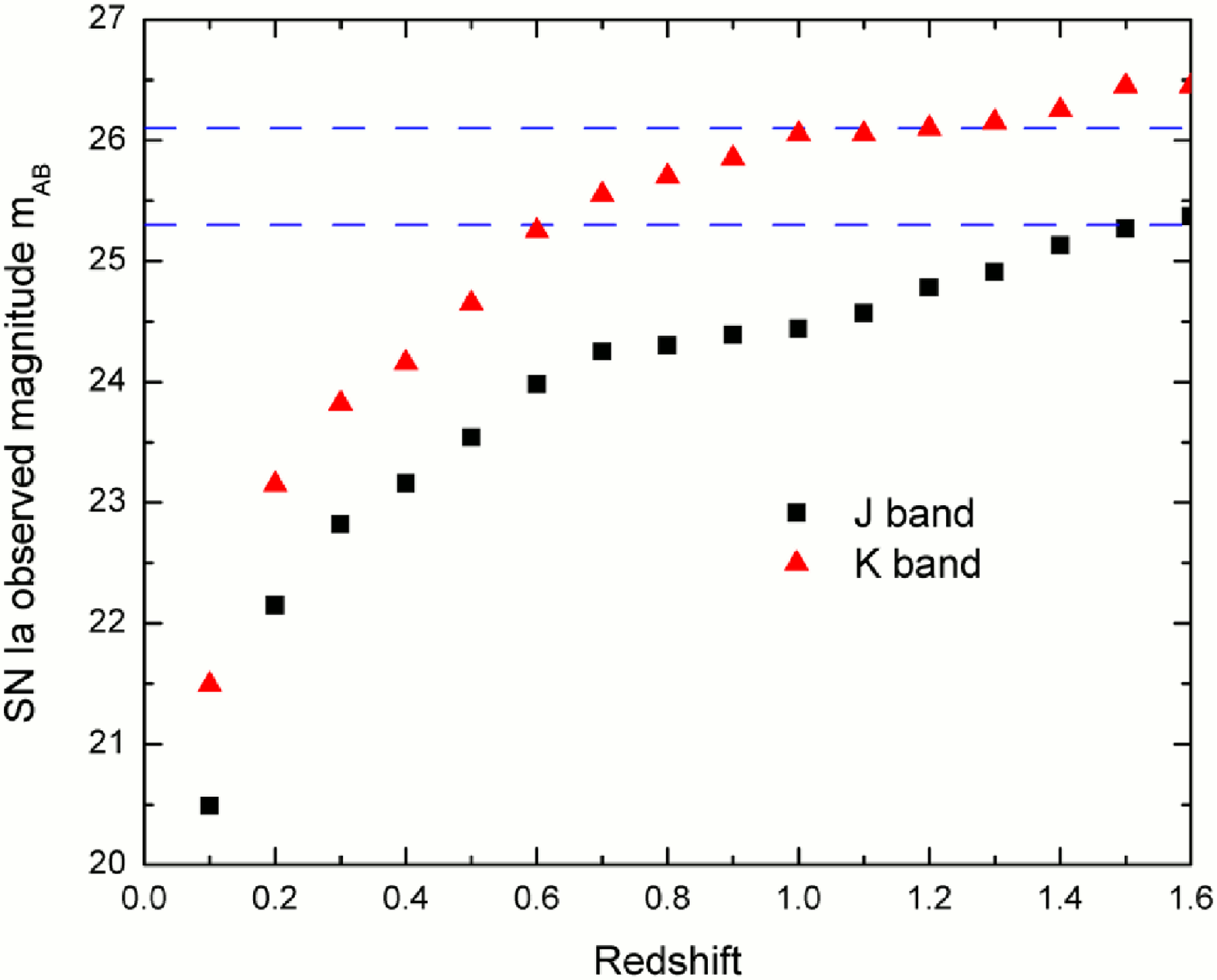}
\caption{The black square data points show the \jb\ observed magnitude within 10~days of
peak brightness for Type Ia supernovae as a function of redshift given by
\citet{Tonry_e_03}. The red triangular data points show an estimate of the \kb\ observed
magnitude based on extrapolations of the \citet{Tonry_e_03} data using the infrared
rest-frame absolute peak brightness given by \citet{Krisciunas_e_04b}. The horizontal
dashed lines show the PILOT PNIRC detection limit for 1~hour and 4~hour integrations at
\kd-band. The \jb\ detection limit is 0.3 magnitudes fainter.}\label{SN_JK}
\end{center}
\end{figure}

\begin{figure}[h]
\begin{center}
\includegraphics[width=7.5cm]{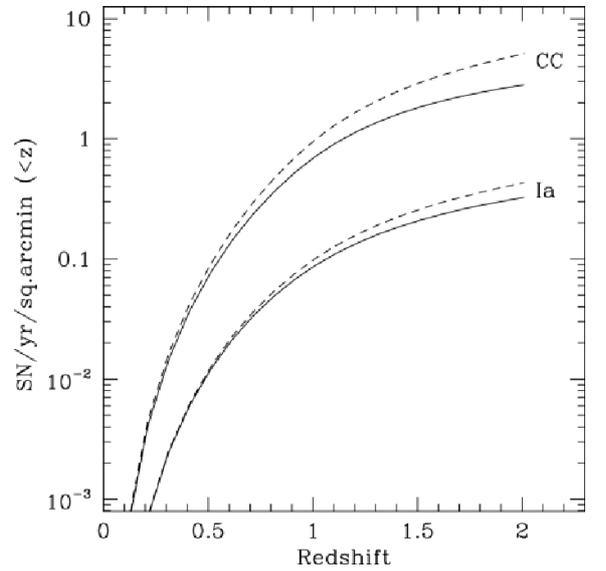}
\caption{Predicted cumulative number of core collapse (CC) and Ia SNa as a function of
redshift. From \citet{Manucci_e_07}.}\label{SN_rate}
\end{center}
\end{figure}

Figure~\ref{SN_curve} shows that SNe stay close to the maximum of light (i.e.,
$\pm$0.25~mag) for about 20~days. The SN detection limits for PILOT as a function of
redshift within this timeframe are shown in Figure~\ref{SN_JK} for \jb\ and \kb\
observations. This plot shows that the optimum near-infrared wavelength to search for
relatively high-redshift ($z > 1$) SNe is \jb. A 1 hour exposure at \jb\ should detect
(at 5~$\sigma$) SN~Ia out to a redshift of $z\approx1.4$. Based on the cumulative
distribution surface density of SN~Ia shown in Figure~\ref{SN_rate}, a dedicated \jb\
search with PNIRC monitoring a \si6~\degsq\ region (200~fields) of sky with 1~hour
exposures and a 10~day cadence would yield \si100~SN~Ia in the range $0.6 < z < 1.0$, and
\si200~SN~Ia in the range $1.0 < z < 1.4$ per season. The majority of these SNe could be
followed-up in \emph{H}- and \emph{K}-bands with \si4~hour exposures, resulting in only a
small reduction in the total expected number of detections. Longer exposures at \emph{L}-
and \emph{M}-bands could also be obtained for a subset of these SN.

While such a dedicated project would yield a significant number of SN Type Ia infrared
light curves, similar detection rates should be possible with, for example, VISTA
operating at \jb. The sensitivity of VISTA at longer wavelengths, however, is not high
enough to allow follow-up of large numbers of detections at \emph{H} and \emph{K}. In
principle, the longer wavelength follow-up could be achieved on an 8~m class telescope
with an adaptive optics system, although this still represents a large observing program
for such a telescope (i.e., several thousand hours).

As opposed to a dedicated search program, a more efficient use of observing time would be
to utilise the data collected from other proposed PILOT wide-field infrared survey
projects. For example, the PISN search (Section~2.1) and the high-redshift galaxy survey
(Section~3.1) will both observe large regions of the sky at \kd. With an appropriate
cadence these searches would be expected to find, for example, \si60~SN~Ia per season out
to $z\approx0.6$ for a shallow survey of $\mab = 25.3$ per field, or \si60~SN~Ia per
season in the range $1.0 < z < 1.6$ for a deeper survey to a depth of $\mab = 26.7$ per
field. Additionally, the weak-lensing survey in the optical will cover a very wide region
of sky (\si4000~\degsq). If this survey were built up from a series of shallow repeat
visits, thousands of low-redshift ($z < 0.5$) SN~Ia would be discovered. SNe detected via
any of these survey projects should be followed up throughout the near-infrared with the
PNIRC instrument. In addition to Type Ia supernovae, core collapse supernovae would also
be found in such searches. The brightness of CC SN is typically 2--3 magnitudes fainter
than Type Ia SN \citep[e.g.,][]{Hamuy_03}, and thus they would be difficult to detect at
high redshifts. However, they occur in much greater number densities, as shown in
Figure~\ref{SN_rate}, and thus should be detectable in large numbers at redshifts below
$z \approx 0.2$.

The SN project proposed here takes advantage of the wide field-of-view, good temporal
coverage, high angular resolution, and high infrared sensitivity of the PILOT telescope.
These characteristics provide a unique opportunity to build an unbiased sample of
infrared light curves of SN~Ia up to a redshift of about $z\approx1.4$. While there are
several current supernovae search programs, e.g., the Canada-France-Hawaii Telescope
Legacy Survey project \citep{Astier_e_06}, these are primarily devoted to finding large
numbers of SNe with multi-colour optical photometry. With a dedicated search program,
PILOT can play an important role in removing (or confirming) concerns that dust may
affect the validity or uncertainty of current constraints placed on cosmological
parameters based on optical data. In addition, the PILOT sample of SN~Ia will form a
subset of the thousands of high-redshift infrared detections possible with the proposed
SNAP mission. This will thus be an important precursor experiment allowing the refinement
of experimental techniques and the control and understanding of systematic effects.

%%%%%%%%%%%%%%%%%%%%%%%%%%%%%%%%%%%%%%%%%%%%%%%%%%%%%%%%%%%%%%%
\section{Conclusion}

The detection and characterisation of the signatures of the first stars to form in the
Universe is observationally very challenging; the PILOT discovery space, however, is well
suited to such science. With a wide-field high-sensitivity near-infrared camera, PILOT is
likely to be one of only two facilities (the other being JWST) capable of detecting
statistically significance numbers of pair-instability supernovae at high redshift. These
objects should provide a rare view of the physical conditions in the early Universe. No
confirmed detection of such an object has yet been made, however, and there are large
uncertainties as to their expected characteristics. The observational parameter space
explored by PILOT and JWST are generally quite distinct, allowing separate regions of the
potential parameter space (i.e., the expected brightness and rarity) for PISN to be
examined. The high sensitivity in the thermal infrared and the large wavelength coverage
of the PILOT infrared instrument suite, combined with the potential for high cadence
observations, are ideally matched to a search for high-redshift gamma-ray burst
afterglows. A dedicated program of high-energy satellite-alert follow-up would be
expected to find a number of very high-redshift gamma-ray burst afterglows each season
and to probe the rate of high-redshift gamma-ray burst orphan-afterglows to a lower
number density than possible with any other facility.

Two projects have been proposed here that demonstrate the potential for PILOT to
investigate the formation and evolution processes of distant galaxies and galaxy
clusters. The first project takes advantage of PILOT's wide field-of-view and high
sensitivity in the near-infrared. These factors enable deep/wide-area near-infrared
surveys to be accomplished that are much deeper and have a higher spatial resolution that
possible with other ground-based telescopes, and that cover a much wider area of sky than
possible with space-based facilities such as JWST. Such surveys will allow the
characteristics of the the earliest galaxy populations at high redshift to be explored in
detail. The second project utilises the wide field-of-view and high spatial resolution of
PILOT at visible wavelengths. These factors allow the assembly processes of galaxy
clusters at moderate redshift to be investigated with much higher efficiency than
possible with other facilities.

PILOT will be particularly capable at investigating the nature and evolution of
cosmological parameters, such as dark energy and dark matter. The wide field-of-view,
high spatial resolution, and stable PSF of the PILOT optical camera are requirements that
are well-suited to the statistical study of weak-lensing effects on distant galaxies; a
method by which the equation of state of the Universe can be investigated. For this
science, PILOT is competitive in terms of survey speed and will achieve tighter
constraints on cosmological parameters than the proposed LSST weak-lensing survey, which
is well ahead of the capabilities of other ground-based telescopes. PILOT is also
competitive in terms of survey speed with the proposed space-based weak-lensing mission,
DUNE; PILOT will not obtain the same level of constraints but its survey can be
accomplished before DUNE achieves first light. Additionally, the wide field-of-view
combined with a high infrared sensitivity and wide wavelength coverage of the PILOT
near-infrared camera will enable it to detect and obtain infrared light-curves for larger
numbers of Type Ia supernovae at higher redshift than possible with any other current
facility. These data will allow tighter constraints to be placed on the expansion of the
Universe by correcting for the effects of dust extinction and reddening.

\section*{Acknowledgments}
The PILOT Science Case, presented here, was produced as part of the PILOT conceptual
design study, funded through the Australian Department of Education, Science, and
Training through the National Collaborative Research Infrastructure Strategy (NCRIS)
scheme, and the University of New South Wales through the UNSW PILOT Science Office. The
European contribution has been supported by the ARENA network of the European Commission
FP6 under contract RICA26150.

\end{document}